\documentclass[twocolumn,showpacs,preprintnumbers,amsmath,amssymb]{revtex4-1}

\usepackage{amsfonts}
\usepackage{amsmath}
\usepackage{amssymb}
\usepackage{graphicx}
\usepackage[usenames,dvipsnames]{color}
\usepackage{pifont}
\usepackage{dcolumn}
\usepackage{bm}

\begin{document}


\title{Analytical maximum-likelihood method to detect patterns in real networks}

\author{Tiziano Squartini}
\affiliation{Dipartimento di Fisica, Universit\`a di Siena, Via Roma 56, 53100 Siena (Italy)}
\affiliation{Instituut-Lorentz for Theoretical Physics, Leiden Institute of Physics, University of Leiden, Niels Bohrweg 2, 2333 CA Leiden (The Netherlands)}
\author{Diego Garlaschelli}%
\affiliation{Instituut-Lorentz for Theoretical Physics, Leiden Institute of Physics, University of Leiden, Niels Bohrweg 2, 2333 CA Leiden (The Netherlands)}
%


\date{\today}

\begin{abstract}
In order to detect patterns in real networks, randomized graph ensembles that preserve only part of the topology of an observed network are systematically used as fundamental null models. However, their generation is still problematic. The existing approaches are either computationally demanding and beyond analytic control, or analytically accessible but highly approximate.
Here we propose a solution to this long-standing problem by introducing a fast method that allows to obtain expectation values and standard deviations of any topological property analytically, for any binary, weighted, directed or undirected network.
Remarkably, the time required to obtain the expectation value of any property analytically across the entire graph ensemble is as short as that required to compute the same property using the adjacency matrix of the single, original network.
Our method reveals that the null behavior of various correlation properties is different from what previously believed, and highly sensitive to the particular network considered.
Moreover, our approach shows that important structural properties (such as the modularity used in community detection problems) are currently based on incorrect expressions, and provides the exact quantities that should replace them.
\end{abstract}

\pacs{Valid PACS appear here}
\maketitle

\section{Introduction}
Detecting relevant patterns in real networks, a fundamental problem for many research fields \cite{barabba,siam,guidosbook}, relies upon the possibility to distinguish the properties explained by the presence of simple constraints from more complex and nontrivial structural features.
For this reason, statistical ensembles of graphs with specified constraints, and otherwise completely random, have been introduced and systematically used as a reference to identify non-random patterns in a real network \cite{MS,MSZ,motifs,generating,chung_lu,newman_origin,
katanza,mygrandcanonical,amaral_foodwebmotifs,
guimera,newman_expo,weightedconfiguration,
serrano_weighted,serrano2,vespy_weighted,
colizza2,manna,ginestra_entropy,mybosefermi,myWRG,milo,uniform,swaps}. Such ensembles serve also as powerful models to study dynamical processes on networks displaying only a set of desired properties, and allow to highlight the dynamical effect of each property separately. The simplest and most important ensembles specify only local constraints. For unweighted networks, this amounts to specify the degree $k_i$ (number of incident edges) of each vertex ($i=1\dots N$ where $N$ is the total number of vertices), and results in the so-called configuration model \cite{MS,MSZ,generating}.
In the weighted case, the corresponding constraint is obtained by fixing the strength $s_i$ (sum of incident edge weights) of each vertex  \cite{weightedconfiguration,serrano_weighted,serrano2}.
More in general, one could enforce different or additional properties \cite{motifs,mygrandcanonical,guimera,newman_expo,serrano_weighted,colizza2,manna,ginestra_entropy,mybosefermi,myWRG}. 

Unfortunately, as we discuss in detail in what follows, it turns out that even in the simplest case with local constraints, the correct generation of random ensembles corresponding to a particular real world network is problematic. 
Both analytical and computational approaches proposed so far have severe limitations.
Motivated by this, here we propose a new maximum-entropy method that is entirely analytical and does not require the generation of randomized variants of a real network. Our method provides the exact probabilities of occurrence of random graphs with the same (average) constraints as the real network, from which the expectation values and standard deviations (and in principle the higher moments) of any topological quantity of interest can be calculated mathematically, either exactly or using proper approximations. Due to its analytical character, our method is extremely faster than all the available alternatives. 
Moreover, it can be applied to undirected, directed, binary and weighted networks in a unified way. 
We will illustrate the power of our approach on several real-world networks of different nature and type, by studying a range of topological properties of interest.

\section{Available methods and their limitations\label{sec_limitations}}
We first briefly review the existing problems in the case of binary unweighted networks, which is the most frequently explored situation.
A binary unweighted graph with $N$ vertices is completely specified by a $N\times N$ adjacency matrix $\mathbf{A}$ with entries $a_{ij}=1$ if the vertices $i$ and $j$ are connected, and $a_{ij}=0$ otherwise.
Generally, one is interested in comparing the observed topological properties of a particular real-world network $\mathbf{A}^*$ against the average properties of a randomized family of networks with the same degree sequence $\vec{k}(\mathbf{A}^*)=\{k_i(\mathbf{A}^*)\}$, where $k_i(\mathbf{A}^*)=\sum_j a_{ij}^*$ is the degree (number of connections) of vertex $i$ in the network $\mathbf{A}^*$.
The ensemble of binary undirected networks with specified degree sequence is known as the \emph{configuration model} (CM) \cite{MS,MSZ,generating} and is currently treated in two very different ways: 
computationally, by explicitly generating many random networks with the desired degree sequence and averaging the quantities of interest across the randomized networks \cite{MS,MSZ}, or analytically, by using approximations that allow to directly estimate the average of topological properties as a function of the enforced degree sequence, without actually measuring them on any network \cite{generating,chung_lu}. 
Currently, both approaches suffer from severe limitations.

\begin{figure}[t!]
\begin{center}
\includegraphics[width=.48\textwidth]{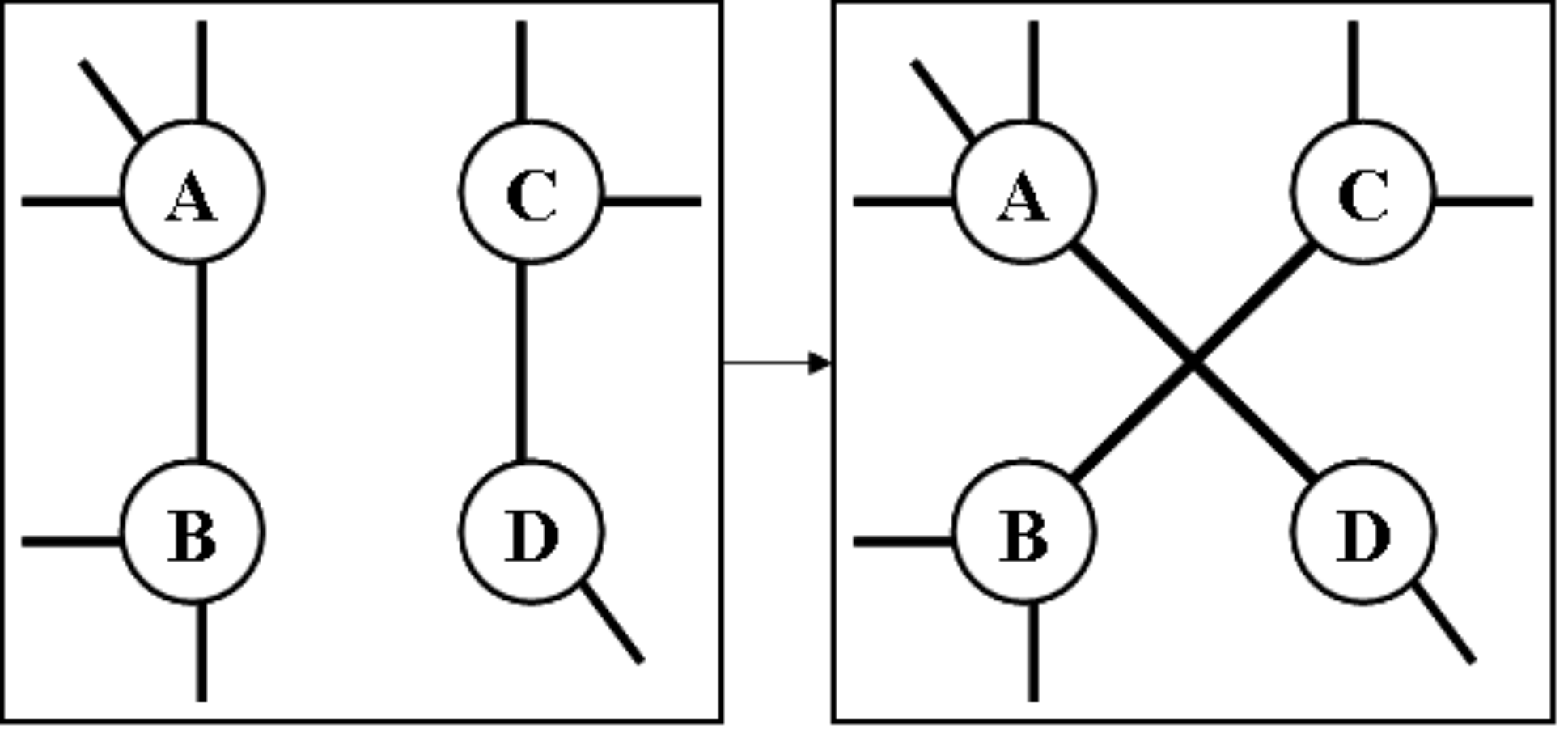}
\end{center}
\caption{An illustration of the local rewiring algorithm whose iteration allows to computationally explore the microcanonical configuration model.\label{fig_LRA}}
\end{figure}

A `bottom-up' computational approach consists in assigning each vertex $i$ a number of `edge stubs' equal to its observed degree $k_i(\mathbf{A}^*)$, and randomly matching pairs of stubs (avoiding self-loops and multiple links) until all degrees reach their desired values (\emph{edge stub reconnection}). However, this procedure is known to get stuck in configurations where vertices requiring additional connections have no more eligible partners \cite{MS,MSZ}. As a consequence, one must implement a `top-down' computational approach where the entire real network $\mathbf{A}^*$ is taken as the initial configuration, and a family of randomized variants is generated by iteratively applying a \emph{local rewiring algorithm} (LRA) where two edges $(A,B)$ and $(C,D)$ are randomly selected and replaced by the two edges $(A,D)$ and $(C,B)$, if the latter are both not already present \cite{MS,MSZ} (see fig.\ref{fig_LRA} for an illustration).
This generates a \emph{microcanonical} ensemble (see the Appendix for a detailed discussion) where all randomized networks have exactly the same degree sequence as the original network, and are sampled with equal probability. This method has been applied to various networks, including the Internet \cite{MSZ}, cellular networks \cite{motifs} and food webs \cite{amaral_foodwebmotifs}, in order to detect higher-order patterns (such as clustering and motifs) not merely due to local constraints. However, this approach is time-consuming since many (a number $R$ much larger than the observed number of links $L$ \cite{MS,milo}, even if not rigorously specified) iterations of the LRA are required to obtain a \emph{single} randomized network, and the entire process must be repeated several times to produce a large number $M$ (again unspecified) of randomized networks, on each of which any topological property $X$ of interest must be measured explicitly and averaged at the end to obtain an estimate for $\langle X\rangle$.  
The computational time required to obtain $\langle X\rangle$ is therefore of the order  $O(M\cdot T_R\cdot R)+O(M\cdot T_X)$, where $T_R$ is the average time required to perform a single successful rewiring step and $T_X$ is that required to compute $X$ on a single network in the randomized set.
Moreover, even if the sufficient statistics of the problem is the degree sequence $\vec{k}(\mathbf{A}^*)$ alone, the above approach requires the entire original network $\mathbf{A}^*$ (or any other network with the same degree sequence, which is however difficult to obtain from scratch due to the problems discussed above) as the starting configuration, thus making use of much more information than required in principle.

By contrast, analytical approaches seek to provide theoretical expressions to directly obtain the ensemble averages of topological properties, without generating the ensemble computationally. Two main approaches exist. One makes use of generating functions for the relevant probability distributions. In the case we are discussing here, the key quantity is the generating function $g(z)=\sum_k z^k P(k)$ of the degree distribution \cite{generating}.
Unfortunately, this method assumes that the network is infinite and locally tree-like (even if in some cases this approximation turns out to perform unexpectedly well even beyond its formal range of applicability \cite{mason}), and is thus inappropriate if the size of the network is small and if the input degree distribution can only be realized by dense and/or clustered networks.  In this approach, clustered or dense networks can only be generated by imposing additional constraints besides the degree sequence, such as the number of triangles attached to vertices \cite{newman_clusteredmodel}, thus leading to a different ensemble which is not the one we are seeking to characterize. 

A different approach looks for an analytical expression for the probability $p_{ij}$ that the vertices $i$ and $j$ are connected in the randomized ensemble  \cite{chung_lu}. Due to its probabilistic nature, this approach generates a \emph{(grand)canonical} ensemble where even graphs violating the constraint are present and assigned different probabilities. In such a case, the constraints are realized on average, i.e. the expectation value $\langle X\rangle$ of any specified property $X$ is fixed exactly (see Appendix). While this approach is indeed very fast in providing averages of the desired properties, it has been shown \cite{newman_origin} that it makes use of a highly approximate expression for $p_{ij}$, valid only when the original network is sparse and/or the degree distribution is not too broad. This expression is
\begin{equation}
p_{ij}=\frac{k_i (\mathbf{A}^*)k_j(\mathbf{A}^*)}{2L^*}
\label{eq_cl}
\end{equation}
where $L^*\equiv L(\mathbf{A}^*)=\sum_i k_i(\mathbf{A}^*)/2=\sum_{i<j}a_{ij}^*$ is the total number of links.
While the expected degree $\langle k_i\rangle=\sum_j p_{ij}$ generated by the above formula coincides with the desired degree $k_i(\mathbf{A}^*)$, the probability $p_{ij}$ may exceed $1$ for pairs of highly connected nodes such that $k_i(\mathbf{A}^*) k_j(\mathbf{A}^*)>2L(\mathbf{A}^*)$. 
In general, only if the degree sequence is such that 
\begin{equation}
k_i(\mathbf{A}^*)<\sqrt{2L(\mathbf{A}^*)}=\sqrt{\sum_j k_j(\mathbf{A}^*)}\quad\forall i
\label{eq_condition}
\end{equation}
then using eq.(\ref{eq_cl}) on the real network $\mathbf{A}^*$ will not lead to the above problem.
While the above condition is typically obeyed by networks with narrow degree distribution such as the Erd\H{o}s-R\'enyi random graph, it is generally violated by scale-free networks displaying a power-law degree distribution $P(k)\sim k^{-\gamma}$, and this violation becomes stronger and stronger as the density of the network increases. 
In particular, it is possible to show that in order to ensure eq.(\ref{eq_condition}) the maximum degree $k_{max}$ in the network should not exceed the so-called \emph{structural cut-off} $k_{c}\sim N^{1/2}$ \cite{cutoff}. This is particularly evident for dense networks where the average degree $\bar{k}=\sum_i k_i/N=2L/N$ remains constant as $N$ increases, so that eq.(\ref{eq_condition}) remains valid only if $k_{max}<\sqrt{2L}\sim N^{1/2}$.
By contrast, extreme value theory shows that in networks with degree distribution $P(k)\sim k^{-\gamma}$ the maximum degree scales as $k_{max}\sim N^{1/(\gamma-1)}$, so that if $\gamma<3$ (as observed in most real-world scale-free networks) then $k_{max}>N^{1/2}$ which exceeds $k_{c}$. 
The meaning of $p_{ij}$ being larger than $1$ for some pairs of vertices in eq.(\ref{eq_cl}) is that, in order to actually realize the degree sequence of the real network $\mathbf{A}^*$, one must let $i$ and $j$ be connected by more than one undirected edge. 
Also, since the desired equality $\langle k_i\rangle=k_i(\mathbf{A}^*)$ is only ensured if one lets the sum in $\sum_j p_{ij}=\langle k_{i}\rangle$ run over all vertices \emph{including $i$ itself}, one must allow the presence of self-loops in the randomized networks.
Thus, even if this is not evident at a first glance, the ensemble generated by eq.(\ref{eq_cl}) does not only contain binary and loop-less undirected graphs and is thus not a proper null model for an empirical binary loop-less network $\mathbf{A}^*$ with degree sequence $\vec{k}(\mathbf{A}^*)$ violating eq.(\ref{eq_condition}), as is typically the case for real-world networks with broad degree distributions.

An elegant proof that the correct ensemble probability $p_{ij}$ for loop-less graphs with no multiple connections differs from eq.(\ref{eq_cl}) has been proposed \cite{newman_origin} and re-derived within the framework of maximum-entropy graph ensembles \cite{newman_expo}.
We shall exploit this result to obtain an exact method later on. We will also show that in real networks the deviation is stronger than expected, and affects sparse networks as well.
An independent proof of the inadequacy of eq.(\ref{eq_cl}) is that it does not generate the graph $\mathbf{A}^*$ with maximum likelihood \cite{mylikelihood}. This can be confirmed by treating $L$ as a free parameter and look for its value $L^{ML}$ that maximizes the probability to obtain $\mathbf{A}^*$. One finds that $L^{ML}\ne L(\mathbf{A}^*)$, which implies that under the maximum likelihood choice $\langle k_i\rangle \ne k_i(\mathbf{A}^*)$ and $\langle L\rangle \ne L(\mathbf{A}^*)$, violating the desired constraint on the degree sequence and the implied one on the number of links \cite{mylikelihood}.
This shows that the functional form of $p_{ij}$ in eq.(\ref{eq_cl}) is intrinsically problematic and does not give highest likelihood to $\mathbf{A}^*$ and to all other graphs with the same degree sequence as $\mathbf{A}^*$.

Therefore, while the available analytical methods are useful to characterise artificially generated networks with special properties, they cannot be used to correctly randomise any real-world network which is either small, clustered, or dense.
Unfortunately, the above limitations are generally ignored, and eq.(\ref{eq_cl}) is frequently used beyond its limits of applicability to estimate connection probabilities. 
Moreover, as we note later on, it is also used as a key ingredient in order to define important structural properties which implicitly rely on a comparison against the CM.
Analogous problems exist in the analysis of directed and/or weighted networks. We will consider each of these cases separately in what follows.

\section{A fast and analytical method}
The above discussion highlights that no method developed so far succeeds in obtaining randomized properties of a particular real-world network such that two requests are met simultaneously:
\emph{i)} the method is general and works for any network, even if displaying small size, high link density, and large clustering; \emph{ii)} expected values across the ensemble can be computed analytically, without sampling the configuration space explicitly. The need to resort to the LRA as the only statistically correct method available, which however requires the artificial generation of many randomized networks, makes the general problem very complicated and all its applications time consuming.

In this paper we propose a solution to this long-standing problem. 
We develop an approach that combines exact expressions for the occurrence probabilities of graphs in maximum-entropy ensembles with given constraints \cite{newman_origin,mygrandcanonical,newman_expo,
ginestra_entropy,mybosefermi,myWRG} with more recent results about the application of the Maximum Likelihood principle to graph ensembles \cite{mylikelihood}.
In the Appendix we describe our method in great detail. 
We start with a general discussion which is formally valid for any constraint, and then consider explicitly the application to real networks where a set of local constraints is enforced. We consider the cases of binary, weighted, directed and undirected networks separately. We show that in all these cases the enforcement of local constraints always leads to exact probabilities that can be easily obtained analytically. Then we also consider an extension to non-local constraints which can still be dealt with analytically. Finally, we compare our (grand)canonical method with the corresponding microcanonical ensemble generated computationally as in the LRA. 

As we show, in all the cases of interest a choice of constraints leads to a specific set of coupled nonlinear equations to be solved. 
In such equations, the observed values of the enforced topological properties (e.g. the degree sequence) determine the values of an equal number of `hidden' parameters in such a way that the real network, or any other network with the same constraints as the real one, is generated with maximum likelihood.
Since only the enforced constraints enter the equations, our method only requires the knowledge of the sufficient statistics of the problem and not of the whole topology, restoring a desirable feature of randomization algorithms. 
Solving the maximum-likelihood equations only takes a computational time $T_E$ which is negligible if compared to the time required to measure any nontrivial topological property, and entirely replaces the  artificial generation of many randomized variants of the original network. 

Once the parameters solving the equations are found, they can be directly used to obtain the expectation value $\langle X\rangle$ and standard deviation $\sigma[X]$ of any topological property $X$ of interest analytically.
When useful, this also allows to obtain a $z$-score representing the number of standard deviations by which the randomized value $\langle X\rangle$ differs form the observed value $X(\mathbf{A}^*)$.
The possibility to obtain the standard deviations and/or $z$-scores is very important, because it allows to assess which topological properties $X$ are consistent with their randomized value $\langle X\rangle$ within a statistical error, and which deviate significantly from the null expectation. In the former case, one can conclude that the enforced constraints completely explain the higher-order property $X$. In the latter case, the observed property cannot be traced back to the constraints, and therefore requires additional explanations or generating mechanisms besides those required in order to explain the constraints themselves (it should be noted, however, that $z$-scores can be unambiguously interpreted only if the property $X$ is normally distributed, and this is generally not the case; nonetheless they still carry information about the discrepancy between observations and the null model).

Importantly, the time required to compute the expectation value $\langle X\rangle$ of a given property $X$ analytically (formally corresponding to an average over a huge number of randomized configurations) is the same as the time $T_X$ required to compute the same property on the single original network. 
Therefore our method takes only a total time $O(T_E+T_X)$ to obtain $\langle X\rangle$ exactly, which is incredibly shorter than the aforementioned time $O(M\cdot T_R\cdot R)+O(M\cdot T_X)$ required by the LRA to obtain $\langle X\rangle$ approximately.
Importantly, $T_E$ is independent of the complexity of the topological property $X$ to measure, which means that for complicated properties $O(T_E+T_X)=O(T_X)$.
Therefore for any topological property $X$ which can be measured in a large but still reasonable time $O(T_X)$ on the real network, the computation of its expectation value $\langle X\rangle$ will require the same time $O(T_X)$. 
If the time required in order to obtain $\langle X\rangle$ is too large, it is because the time required to measure $X$ is too large as well. In other words, the property $X$ is too complicated to be computed on the real network itself.
In such a case, the problem is not due to the method, but to a demanding choice of $X$ for that particular network. Note that we are assuming that the topological properties of the real network are computed using the full adjacency matrix. This is the worst-case scenario, since in many cases (especially for sparse networks) it is enough to use reduced information such as the list of existing links. For instance, the time to measure the clustering coefficient can be significantly shorter, using optimized algorithms, on a sparse network than on a generic network of the same size (an in this case it will also be shorter than the time required to compute its randomized value across our ensemble). However, our interest is precisely to focus on the (worst) general case (e.g. dense and very dense networks), because it is in this case that other approaches fail (such as eq. \ref{eq_cl}), or become extremely time consuming (such as the LRA, which takes longer for denser graphs). 

\section{Results}
We now show the application of our method to real networks of various type, by considering several topological properties and their randomized counterparts. 

\subsection{Binary undirected networks}
We start with the simplest case of binary undirected networks.
One of the most important topological properties of a binary network is the correlation between the degrees of adjacent nodes, which has been shown to dramatically affect various structural and dynamical features \cite{siam}. These correlations can be measured by the average nearest neighbour degree (ANND), which on the real network $\mathbf{A}^*$ is defined as
\begin{equation}
k^{nn}_i(\mathbf{A}^*)\equiv
\frac{\sum_{j\ne i}\sum_{k\ne j} a^*_{ij}a^*_{jk}}
{\sum_{j\ne i} a^*_{ij}}
\label{eq_knn}
\end{equation}
While the degree is a first-order property which only depends on the number of links (topological paths of length one) entering a vertex, the ANND is a second-order property contributed by paths of length $2$ (i.e. the terms $a^*_{ij}a^*_{jk}$). 
Similarly, a third-order (i.e. involving paths of length 3) property is the clustering coefficient $c_i$, which represents the fraction of pairs of neighbours of vertex $i$ which are mutually connected:
\begin{equation}
c_i(\mathbf{A}^*)\equiv
\frac{\sum_{j\ne i}\sum_{k\ne i,j}a^*_{ij}a^*_{jk}a^*_{ki}}
{\sum_{j\ne i}\sum_{k\ne i,j}a^*_{ij}a^*_{ki}}
\label{eq_c}
\end{equation}

\begin{figure}
\begin{center}
\includegraphics[width=.48\textwidth]{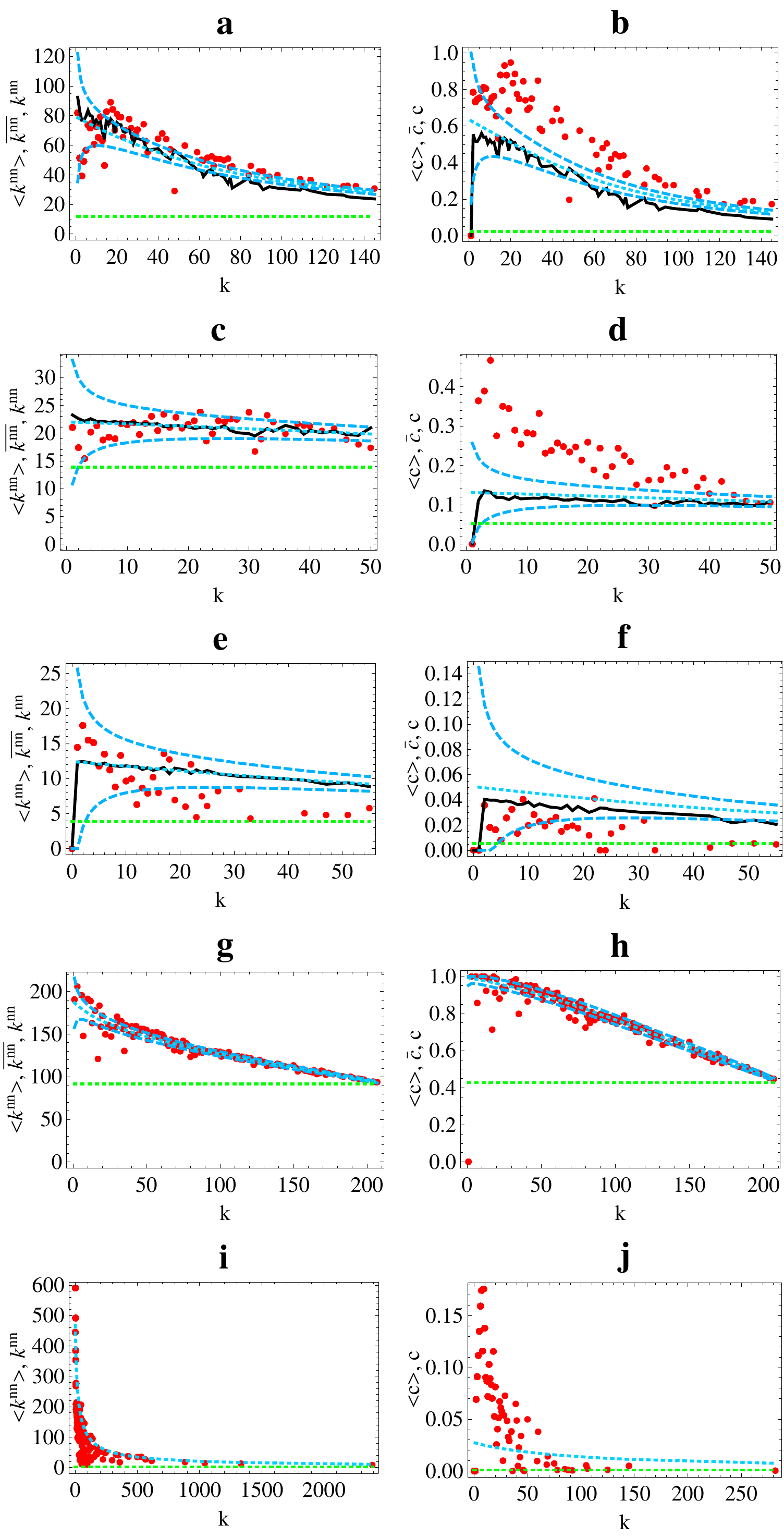}
\end{center}
\caption{Application of our method to binary undirected networks. The red points are the empirical data, the black solid curves are averages over the configuration model obtained using the local rewiring algorithm \cite{MS,MSZ}, and the blue dashed curves are the analytical expectations ($\pm$ one standard deviation) obtained using our method. The green curves are the flat expectations under the Erd\H{o}s-R\'enyi random graph model, and highlight the average level of correlation in the random case. The panels report
$k^{nn}_i$ versus $k_i$ (left) and $c_i$ versus $k_i$ (right) for: a) and b) the network of the largest US airports ($N=500$) \cite{airportdata}, c) and d) the synaptic network of \emph{Caenorhabditis elegans} ($N=264$) \cite{neural}, e) and f) the protein-protein interaction network of \emph{Helicobacter pylori} ($N=732$) \cite{ppidata}, g) and h) the network of liquidity reserves exchanges between Italian banks in 1999 \cite{interbank} ($N=215$), i) 
the Internet at the AS level ($N=11.174$) \cite{internetdata} and j) the protein-protein interaction network of \emph{Saccharomices cerevisiae} ($N=4.142$) \cite{ppidata}. The last two networks are randomized using only our method, as the local rewiring algorithm would require much more time given the large number of edges.
\label{fig_undirected}}
\end{figure}
As we mentioned, it is always important to assess whether in a particular real network higher-order properties arise merely as a consequence of low-level constraints or whether they signal additional structural patterns. 
In particular, comparing the real network $\mathbf{A}^*$ with the CM (which provides an ensemble of random networks having, on average, the same degree sequence $\vec{k}(\mathbf{A}^*)$ as $\mathbf{A}^*$) allows to assess whether longer topological paths and the structural properties involving them are simply a random concatenation of the individual links enforced by the degree sequence, or whether they are irreducible to first-order constraints.
As we discuss in detail in the Appendix, our method can solve this problem by making use of an auxiliary $N$-dimensional vector $\vec{x}=\{x_1\dots x_N\}$ of parameters.
In particular, one must look for the particular values $\vec{x}^*$ that solve the following set of $N$ coupled nonlinear equations:
\begin{equation}
\sum_{j\ne i}\frac{x^*_ix^*_j}{1+x^*_ix^*_j}=k_i(\mathbf{A}^*) \qquad \forall i
\label{eq_k}
\end{equation}
where $k_i(\mathbf{A}^*)$ is the observed degree of vertex $i$ in the real network $\mathbf{A}^*$.
Once the parameter values are found, they allow to obtain analytically the expectation value $\langle X\rangle^*$ of any topological property $X$ across the desired ensemble. This simply amounts to replace the adjacency matrix entry $a^*_{ij}$ appearing in the definition of $X(\mathbf{A}^*)$ with its expectation value 
\begin{equation}
p_{ij}^*=\langle a_{ij}\rangle^*=\frac{x_i^*x_j^*}{1+x_i^*x_j^*}
\label{eq_acorrect}
\end{equation}
which represents the correct expression that should be used in place of eq.(\ref{eq_cl}). 
Similarly, it is possible to obtain the standard deviation $\sigma^*[X]$ analytically in terms of $\vec{x}^*$ (see Appendix). 

In fig.\ref{fig_undirected} we show an application of our method on the network of the 500 largest US airports \cite{airportdata}, a synaptic network \cite{neural}, two protein interaction networks \cite{ppidata}, an interbank network \cite{interbank} and the Internet at the Autonomous Systems level \cite{internetdata}. These are among the most studied networks of this type.
We compare the correlation structure of the original networks, as measured by the dependence of $k^{nn}_i(\mathbf{A}^*)$ and $c_i(\mathbf{A}^*)$ on $k_i(\mathbf{A}^*)$, with the expected values $\langle k^{nn}_i\rangle^*$ and $\langle c_i\rangle^*$ obtained analytically using our method.
Note that we are averaging the values of $k^{nn}_i(\mathbf{A}^*)$ and $c_i(\mathbf{A}^*)$ over all vertices with the same degree: this makes our comparison with the values $\langle k^{nn}_i\rangle^*$ and $\langle c_i\rangle^*$ consistent, since both real and randomized quantities can be plotted using the same values $\langle k_i\rangle^*=k_i(\mathbf{A}^*)$ on the abscissa (we use the same strategy in what follows).
We also highlight the region within one standard deviation around the average by plotting the curves $\langle k^{nn}_i\rangle^*\pm \sigma^*[k^{nn}_i]$ and $\langle c_i\rangle^*\pm \sigma^*[c_i]$. 
For the sake of comparison, we also report the average values obtained sampling the microcanonical ensemble with the standard local rewiring algorithm \cite{MS,MSZ}, and the expected values over the ensemble of random graphs with the same number of links (\textit{random graph model}, RG)
As we mentioned, the microcanonical method requires the generation of many randomized variants, many rewirings per variant, and the measurement of $k^{nn}_i$ and $c_i$ on each variant separately, plus a final averaging. By contrast, our method only requires the preliminary estimation of the $\{x^*_i\}$. Then the calculation of $\langle k^{nn}_i\rangle$ and $\langle c_i\rangle$ takes exactly the same time as that of the empirical values. 
As can be seen, the two approaches yield very similar results (in the Appendix we provide a detailed comparison of the two methods). For the two largest networks (the protein interactions in \emph{S. cerevisiae} and the Internet), we only report the expectations obtained using our method, as the microcanonical approach would require too much computing time. 

The above results allow to interpret the effect of the degree sequence on higher-order properties.
Firstly, the trends displayed by the CM are not flat as those expected in the random graph case. This confirms that residual structural correlations, simply due to the enforced constraint, are still present after the rewiring has taken place. The presence of these correlations does not require any additional explanation besides the existence of the constraints themselves. This is very different from the picture one would get by using the (wrong) expectation of eq.(\ref{eq_cl}) which would yield flat trends as well, naively suggesting that correlations can never be traced back to the degree sequence alone.
Secondly, while the trends observed in all the networks considered are always decreasing, they unveil different correlation patterns when compared to the randomized trends.
The real interbank data are almost indistinguishable from the randomized curves, meaning that structural constraints can fully explain the observed behaviour of higher-order network properties.
Instead, in the airport network the randomized curves lie below the real data (except for an opposite trend of $\langle k^{nn}_i\rangle$ for low degrees). This means that the real network is more correlated than the baseline randomized expectation, and indicates that additional mechanisms producing positive correlations must be present on top of structural effects. By contrast, in the \emph{H. pylori}'s protein network the expected curves lie above the real data, suggesting the presence of mechanisms producing negative correlations. The same is true for the correlation structure of the Internet, confirming previous results \cite{MSZ}, while \emph{S. cerevisiae}'s protein network is completely different from its randomized variants. Therefore seemingly similar trends can actually reveal very different types of structural organization. This means that measuring the topological properties alone is uninformative, and makes the comparison between real data and randomized ensembles essential. Thus the possibility to analytically and quickly characterize the latter, which was previously unavailable, is a remarkable advantage of our approach.

\subsection{Directed networks}
We now consider binary directed networks, which are specified by an asymmetric adjacency matrix $\mathbf{A}$. The local constraints are now represented by the joint sequence of out-degrees and in-degrees $\{k^{out}_i,k^{in}_i\}=\{\sum_{j\ne i}a_{ij},\sum_{j\ne i}a_{ij}\}$.
Given a particular real network $\mathbf{A}^*$ and a measured topological property $X(\mathbf{A}^*)$, our method allows to analytically obtain the expectation value $\langle X\rangle^*$ and standard deviation $\sigma^*[X]$ across the ensemble of binary directed graphs with, on average, the same directed degree sequences $\vec{k}^{out}(\mathbf{A}^*)$ and $\vec{k}^{in}(\mathbf{A}^*)$ as $\mathbf{A}^*$ (\textit{directed configuration model}, DCM).
As shown in the Appendix, in this case our method makes use of two $N$-dimensional vectors $\vec{x}$, $\vec{y}$ of auxiliary variables, and requires that these parameters are set to the particular values $\vec{x}^*$, $\vec{y}^*$ that solve the following set of $2N$ coupled nonlinear equations:
\begin{eqnarray}
\sum_{j\ne i}\frac{x^*_iy^*_j}{1+x^*_iy^*_j}&=&k^{out}_i(\mathbf{A}^*)\qquad \forall i\\
\sum_{j\ne i}\frac{x^*_jy^*_i}{1+x^*_jy^*_i}&=&k^{in}_i(\mathbf{A}^*)\qquad \forall i
\label{eq_koutin}
\end{eqnarray}
The quantities $\vec{x}^*$, $\vec{y}^*$ allow to obtain $\langle X\rangle^*$ and $\sigma^*[X]$ analytically and quickly, outperforming the directed version of the LRA \footnote{In the directed version of the local rewiring algorithm, two directed edges $(A,B)$ and $(C,D)$ are randomly selected and replaced with the directed edges $(A,D)$ and $(C,B)$, if the latter are not already present.}. Note that, as in the undirected case, the method only makes use of the sufficient statistics of the problem.

\begin{figure}
\begin{center}
\includegraphics[width=.48\textwidth]{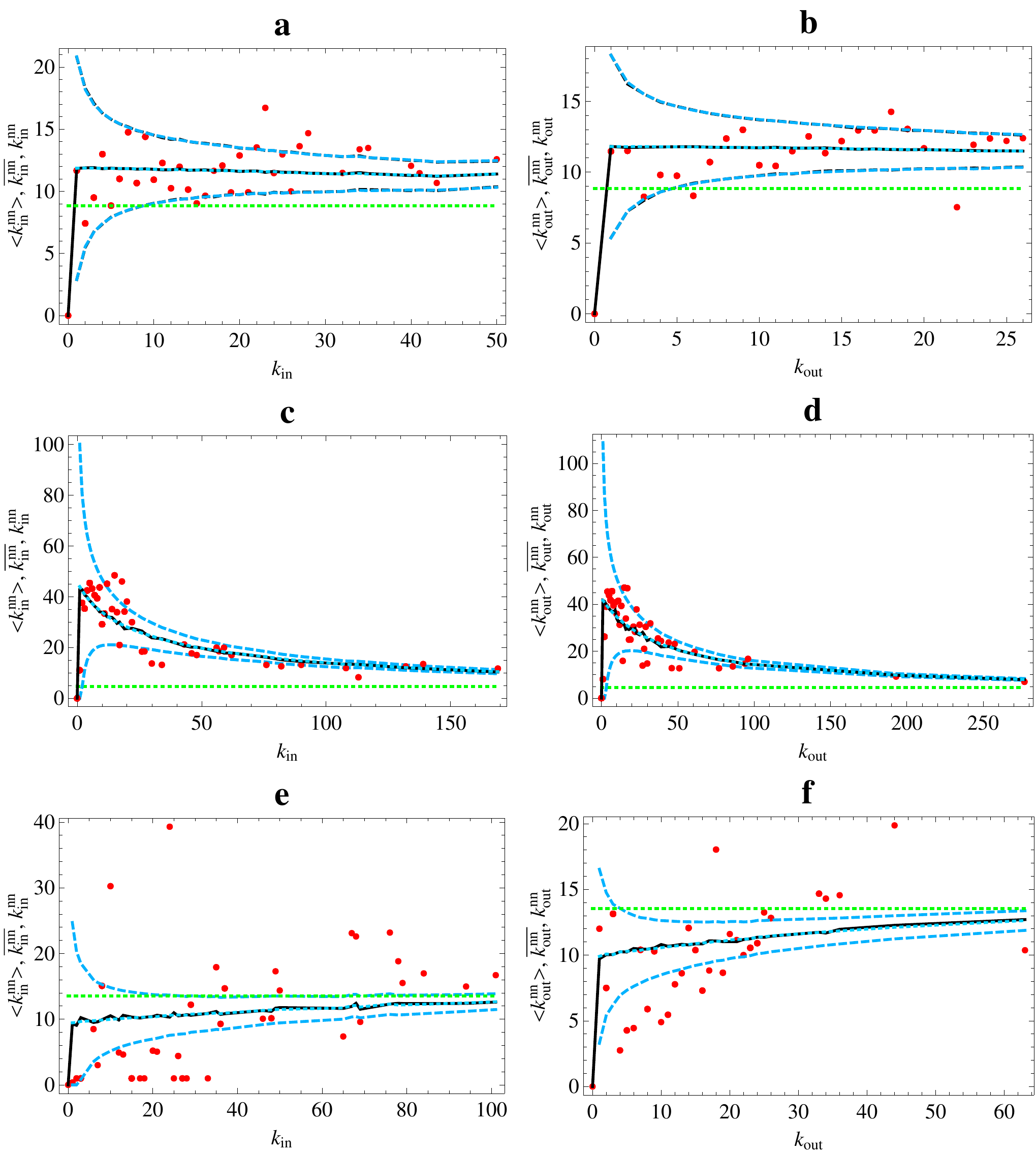}
\end{center}
\caption{Application of our method to directed networks. Red points are the empirical data, the black solid curves are expectations under the directed configuration model using the local rewiring algorithm, and the blue dashed curves are the exact expectations obtained using our method ($\pm$ one standard deviation). 
The green curves are the flat expectations under the directed version of the Erd\H{o}s-R\'enyi random graph model.
The panels report $k^{nn,in}_i$ versus $k^{in}_i$ (left) and $k^{nn,out}_i$ versus $k^{out}_i$ (right) for: a) and b) the directed neural network of \emph{Caenorhabditis elegans} ($N=264$) \cite{neural}, c) and d) the metabolic network of \emph{Escherichia coli} ($N=1078$) \cite{metabolic}, e) and f) the \emph{Little Rock Lake} food web ($N=183$) \cite{littlerock}.
For the \emph{C. elegans} network, we also show the microcanonical standard deviations obtained using the LRA (black dotted curves), which are indistinguishable from the grandcanonical ones.
\label{fig_directed1}}
\end{figure}

We apply our method to various directed networks, by studying the second-order topological properties measured by the outward ANND and the inward ANND, which are defined as two natural generalizations of eq.(\ref{eq_knn}):
\begin{eqnarray}
k^{nn,out}_i(\mathbf{A}^*)&\equiv&
\frac{\sum_{j\ne i}\sum_{k\ne j} a^*_{ij}a^*_{jk}}
{\sum_{j\ne i} a^*_{ij}}\label{eq_knnout}\\
k^{nn,in}_i(\mathbf{A}^*)&\equiv&
\frac{\sum_{j\ne i}\sum_{k\ne j} a^*_{ji}a^*_{kj}}
{\sum_{j\ne i} a^*_{ji}}
\label{eq_knnin}
\end{eqnarray}
In fig.\ref{fig_directed1} we plot the observed values $k^{nn,in}_i(\mathbf{A}^*)$ versus $k^{in}_i(\mathbf{A}^*)$ and $k^{nn,out}_i(\mathbf{A}^*)$ versus $k^{out}_i(\mathbf{A}^*)$, as well as the expectations $\langle k^{nn,in}_i\rangle^*\pm\sigma^*[k^{nn,in}_i]$ and $\langle k^{nn,out}_i\rangle^*\pm\sigma^*[k^{nn,out}_i]$ obtained using our model (see Appendix), for three real directed networks: the neural network of \emph{C. elegans} \cite{neural} (now in its directed version), the metabolic network of \emph{E. coli} \cite{metabolic}, and the \emph{Little Rock Lake} food web \cite{littlerock}.
As before, we also show the microcanonical average obtained using the LRA and the expectation under the \textit{directed random graph model} (DRG) with the same number of links. Again, we find a very good agreement between the two approaches, confirming that our method yields the correct prediction in incredibly shorter time (see Appendix for a discussion about the convergence time of the LRA to our exact results). For the \emph{C. elegans} network (fig. \ref{fig_directed1}a-b), we also show the microcanonical standard deviations, which turn out to be indistinguishable from the grandcanonical ones.
We also confirm that while some networks (\emph{C. elegans} and \emph{E. coli}) are almost consistent with the null model, others (\emph{Little Rock}) deviate significantly.

\begin{figure}[h!]
\begin{center}
\includegraphics[width=.48\textwidth]{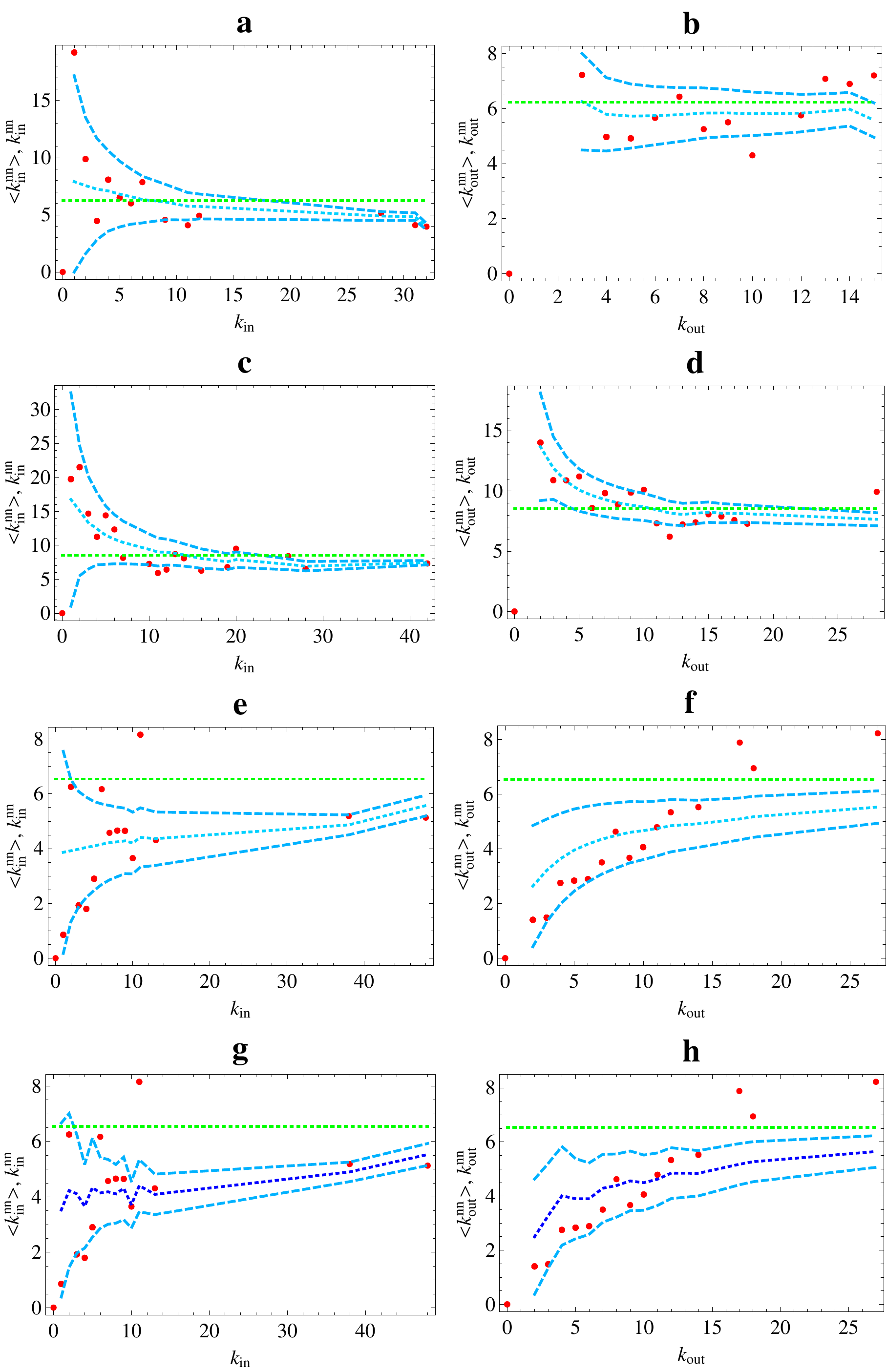}
\end{center}
\caption{Application of our method to small-size directed food webs. Red points are the empirical data and the blue dashed curves are the exact expectations ($\pm$ one standard deviation) under the directed configuration model obtained using our method. 
The green curves are the flat expectations under the directed version of the Erd\H{o}s-R\'enyi random graph model.
The panels report $k^{nn,in}_i$ versus $k^{in}_i$ (left) and $k^{nn,out}_i$ versus $k^{out}_i$ (right) for: a) and b) the Narragansett Bay web ($N=35$) \cite{pajek}, c) and d) the Mondego Estuary web ($N=46$) \cite{pajek}, e) and f) the St. Marks River web ($N=54$) \cite{pajek}. For the latter, in g) and h) we also compare the empirical data with the expectations under the reciprocal configuration model, where also the number of reciprocated links of each vertex is specified.\label{fig_foodwebs}}
\end{figure}

However, the most interesting point for the present analysis is that, while for the undirected networks considered above all randomized trends were decreasing, in this case we find that the three randomized trends behave in totally different ways. 
In the neural network, both $\langle k^{nn,in}_i\rangle^*$ and
$\langle k^{nn,out}_i\rangle^*$ are approximately constant. This means that the baseline behavior for both quantities is flat and uncorrelated (as in the directed random graph, but at a different level). 
By contrast, in the metabolic network the expected curves are decreasing, and thus the ensemble of randomized networks is disassortative as for the undirected graphs considered above.
Finally, in the food web the constraints enforce unusual positive correlations, and the randomized ensemble is even assortative.
Interestingly, while it is expected that random networks with specified degrees display a disassortative behavior \cite{MSZ,newman_origin}, the assortative trend is totally surprising. This is because our method extracts the hidden variables directly from the specific real world network, rather than drawing them from ad hoc distributions. The resulting values can be distributed in a very complicated fashion, invalidating the results obtained under other hypotheses.
To further highlight this important point, we selected three more food webs characterized by a particularly small size (see fig.\ref{fig_foodwebs}). 
Small networks cannot be described by approximating the mass probability function of their topological properties (such as the degree) with a continuous probability density. 
Therefore in this case the difference between the expectations obtained by drawing the $\vec{x}$ and $\vec{y}$ values from analytically tractable continuous distributions and those obtained by solving eqs.(\ref{eq_koutin}) using the empirical degrees is particularly evident.
As we show in fig.\ref{fig_foodwebs} (where for simplicity we omit the comparison with the LRA), we confirm that the (directed) CM can display not only flat or decreasing trends, but also increasing ones. Importantly, in this case all three webs do not deviate dramatically from the null model. This means that while one would be tempted to interpret the three observed trends as signatures of different patterns (zero, negative and positive correlation), actually in all three cases the observed behavior can be roughly replicated by the same mechanism and almost entirely traced back to the degree sequence only. This unexpected result highlights once again that the measured values of any topological property are \emph{per se} entirely uninformative, and can only be interpreted in relation to a null model.

\subsection{Reciprocity and motifs}
So far, in our analysis of directed networks we have considered second-order topological properties. In principle, third-order properties can be studied by introducing directed generalizations of the clustering coefficient \cite{fagiolo_clustering,fink_clustering}. However, there is a proliferation of possible third-order patterns due to the directionality of links. For this reason, a more complete analysis consists in counting (across the entire network) all the possible directed \emph{motifs} \cite{motifs} involving three vertices, and comparing the empirical abundances with the expected ones under the null model. 
As we show in a moment, our method lends itself admirably in such a case. 
Before presenting our results, we note however that directionality makes the possible specifications of the null model proliferate as well. In particular, besides the DCM considered above, a more refined way to randomize directed networks includes the possibility to enforce additional constraints on the reciprocity structure \cite{motifs,mygrandcanonical}. 
In other words, it is possible (and important in many applications \cite{motifs,amaral_foodwebmotifs}) to preserve not only the total numbers $k^{in}_i$ and $k^{out}_i$ of incoming and outgoing links of each vertex, but also the number $k^{\leftrightarrow}_i\equiv \sum_j a_{ij}a_{ji}$ of reciprocated links (pairs of links in both directions) \cite{myreciprocity,vinkosreciprocity}.
This specification is equivalent to enforce, for each vertex $i$, the three quantities \cite{myreciprocity,mygrandcanonical} $k^{\rightarrow}_i\equiv \sum_{j\ne i} a^\rightarrow_{ij}$ 
(number of non-reciprocated outgoing links), $k^{\leftarrow}_i\equiv \sum_{j\ne i} a^\leftarrow_{ij}$ (number of non-reciprocated incoming links) and $k^{\leftrightarrow}_i\equiv \sum_{j\ne i} a^\leftrightarrow_{ij}$ (number of reciprocated links), where $a^\rightarrow_{ij}\equiv a_{ij}(1-a_{ji})$, $a^\leftarrow_{ij}\equiv a_{ji}(1-a_{ij})$ and $a^\leftrightarrow_{ij}\equiv a_{ij}a_{ji}$.

Given a real directed network $\mathbf{A}^*$, we denote the null model with specified joint reciprocal degree sequences $\{k^{\rightarrow}_i(\mathbf{A}^*),
k^{\leftarrow}_i(\mathbf{A}^*),
k^{\leftrightarrow}_i(\mathbf{A}^*)\}$ as the \emph{reciprocal configuration model} (RCM).
This is an example of model with nonlocal (second-order) constraints which can still be treated analytically using our method.
As we show in the Appendix, in this case one must solve the following $3N$ coupled equations:
\begin{eqnarray}
\sum_{j\ne i}\frac{x^*_i y^*_j}{1+x^*_i y^*_j+x^*_j y^*_i+ z_i^* z_j^*}
&=&k^\rightarrow_i(\mathbf{A}^*)\qquad \forall i\\
\sum_{j\ne i}\frac{x^*_j y^*_i}{1+x^*_i y^*_j+x^*_j y^*_i+ z_i^* z_j^*}
&=&k^\leftarrow_i(\mathbf{A}^*)\qquad \forall i\\
\sum_{j\ne i}\frac{z^*_i z^*_j}{1+x^*_i y^*_j+x^*_j y^*_i+ z_i^* z_j^*}
&=&k^\leftrightarrow_i(\mathbf{A}^*)\qquad \forall i
\end{eqnarray}
The expectation value of any topological property, as well as its standard deviation, can now be calculated analytically in terms of the three $N$-dimensional vectors $\vec{x}^*$, $\vec{y}^*$, $\vec{z}^*$.
For instance, in fig.\ref{fig_foodwebs}g-h we repeat the analysis of the directed ANND of the St. Marks River food web, now comparing the observed trend against the RCM. In this case, we find no significant difference with respect to the DCM considered above (fig.\ref{fig_foodwebs}e-f). However, as we now show, the analysis of motifs reveals a dramatic difference between the predictions of the two null models.

\begin{figure}[t!]
\begin{center}
\includegraphics[width=.5\textwidth]{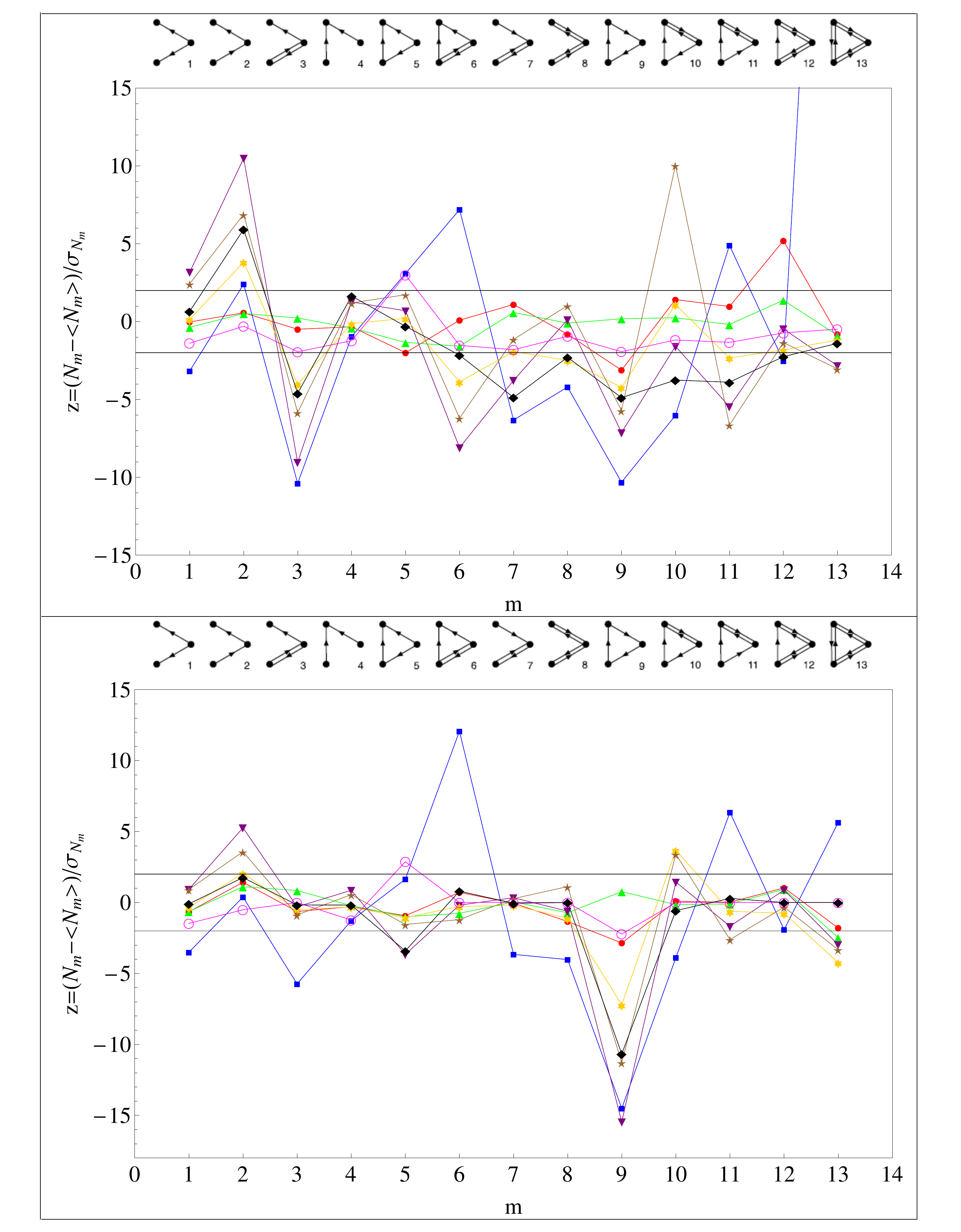}
\end{center}
\caption{Legend: $\textcolor{red}{\bullet}$ - Chesapeake Bay, $\textcolor{blue}{\blacksquare}$ - Little Rock Lake, $\textcolor{green}{\blacktriangle}$ - Maspalomas Lagoon, $\textcolor{Purple}{\blacktriangledown}$ - Florida Bay, $\textcolor{Dandelion}{*}$ - St Marks Seagrass, $\textcolor{Brown}{\star}$ - Everglades Marshes, $\textcolor{CarnationPink}{\circ}$ - Grassland, $\textcolor{black}{\blacklozenge}$ - Ythan Estuary.\\ Application of our method to the analysis of directed \emph{motifs} involving three vertices in 8 real food webs. Top panel: $z$-scores obtained enforcing only the in-degree and out-degree sequences (directed configuration model). Bottom panel: $z$-scores obtained enforcing also the reciprocal degree sequence (reciprocal configuration model).
\label{fig_motifs}}
\end{figure}

If $N_m$ denotes the number of occurrences of a particular motif $m$, our method allows to calculate the expected number $\langle N_m\rangle^*$ and standard deviation $\sigma^*[N_m]$ exactly (see Appendix), and thus to obtain the $z$-score
\begin{equation}
z[N_m]\equiv\frac{N_m(\mathbf{A}^*)-\langle N_m\rangle^*}{\sigma^*[N_m]}
\end{equation}
analytically. This can be done for both the DCM and the RCM.
The value of $z[N_m]$ indicates by how many standard deviations the observed and expected numbers of occurrences of motif $m$ differ. Large values of $z[N_m]$ indicate motifs that are either over- or under-represented under the particular null model considered, and that are therefore not explained by the lower-order constraints enforced.
In fig.\ref{fig_motifs} we show the $z$-scores for all the possible 13 non-isomorphic connected motifs with three vertices in 8 real food webs, for both null models. We also show the two lines $z=\pm 2$ to highlight the region within 2 standard deviations from the model's expectations.
This analysis is similar to that of ref.\cite{amaral_foodwebmotifs}, but is made much simpler by our method which does not require to randomize the webs through a computational algorithm preserving the (reciprocal) degree sequences.
The food webs considered here are from different ecosystems (lagoons, marshes, lakes, bays, estuaries, grasses), with a prevalence of aquatic habitats.
The presence of (intrinsically directed) predator-prey relationships implies that reciprocity is a very important quantity in food webs \cite{amaral_foodwebmotifs}. Thus the RCM should fluctuate less than the DCM. 
Indeed, this is confirmed by our analysis.
The $z$-scores for the motifs $m=2,3,13$ are significantly reduced from the DCM to the RCM.
Also, while the motifs $m=1,6,10,11$ display large values of $z$ with opposite signs across different webs under the DCM, the signs of all statistically surprising motifs (i.e. when $|z|\gtrsim 2$) become consistent with each other under the RCM (except for $m=13$).
As a consequence, under the RCM all networks display a very similar pattern, and the most striking features of real webs become the over-representation of motifs $m=2,10$ (plus $m=6,11,13$ for the Little Rock Lake web) and the under-representation of motifs $m=5,9,13$ (plus $m=3,7,8$ for Little Rock Lake). In particular, the under-representation of motif $m=9$ (the 3-loop) is the most common pattern across all webs, and becomes stronger as the reciprocity of the web increases.
Also note that in a network with no reciprocated links, the number of motifs with at least a pair of reciprocated links is zero. Under the RCM, the expected number of these motifs remains zero. By contrast, their expected number under the DCM is always positive.
Thus we confirm that the upgrade to the RCM is necessary, as its stricter constraints allow to analyze 3-vertices motifs once 2-vertices motifs (i.e. all possible dyadic patterns) are correctly accounted for.
The possibility to treat the RCM analytically using our method is therefore an important step forward.  

\subsection{Weighted networks}
Remarkably, our method works equally well for weighted graphs (where the binary adjacency matrix $\mathbf{A}$ is replaced by a non-negative weight matrix $\mathbf{W}$), thanks to recent analytical results that allow to characterize maximally random weighted networks with specified properties in a way that is completely analogous to their binary counterparts \cite{mybosefermi,myWRG}.
In a particular weighted network $\mathbf{W}^*$, the local constraints are the strength sequence $\{s_i(\mathbf{W}^*)\}=\{\sum_j w^*_{ij}\}$ (undirected case) or the joint out-strength and in-strength sequence $\{s^{out}_i(\mathbf{W}^*),s^{in}_i(\mathbf{W}^*)\}=\{\sum_j w^*_{ij},\sum_j w^*_{ji}\}$ (directed case).
We will only consider undirected weighted networks. The extension to the directed case is straightforward. 
The family of randomized weighted graphs with the same strength sequence as a real weighted network is sometimes denoted as the \emph{weighted configuration model} (WCM) \cite{weightedconfiguration}.
The available microcanonical algorithms regard each link weight as an integer multiple $w$ of a fundamental unit of weight, transform each edge of weight $w$ into $w$ edges of unit weight, and rewire the latter as in the unweighted case, now ensuring that the strength (total number of incoming edges of unit weight) of each vertex is preserved. 
This means replacing a list of $L^*\le N(N-1)/2$ weighted links, summing up to a total weight $W^*=\sum_{i<j}w^*_{ij}$, with $W^*\gg N(N-1)/2$ unweighed links. As real networks have broadly distributed weights summing up to a large $W^*$, this procedure becomes very time consuming as incredibly many rewiring steps per randomized variant must be performed.
As for the binary case, an alternative procedure makes use of a naive theoretical expectation \cite{weightedconfiguration,newman_weighted} for the expected weight of a link in the WCM, in analogy with eq.(\ref{eq_cl}):

\begin{equation}
\langle w_{ij}\rangle=\frac{s_i(\mathbf{W}^*)s_j(\mathbf{W}^*)}{2W^*}
\label{eq_clw}
\end{equation}

However, the above expression has been shown to have as many limitations as its binary counterpart, and to be incorrect \cite{mybosefermi}. 

By contrast, as we show in the Appendix, our method allows to treat the WCM analytically as in the unweighted case. 
Note that choosing the unit of weight in the WCM (before performing the randomization) is in principle arbitrary, but the resulting ensemble will be different for different choices. This issue of granularity is an open problem that deserves future investigations. Our grandcanonical alternative to the WCM is not aimed at fixing the problem, but at providing, for a given choice of the weight unit in the microcanonical ensemble, the corresponding grandcanonical expectation.

Given a real weighted undirected network $\mathbf{W}^*$, our method proceeds by finding the particular values $\{x^*_i\}$ solving the $N$ coupled equations
\begin{equation}
\sum_{j\ne i}\frac{x^*_ix^*_j}{1-x^*_ix^*_j}=s_i(\mathbf{W}^*)\qquad \forall i
\label{eq_s}
\end{equation}
Note the difference of sign with respect to eq.(\ref{eq_k}).
As in the binary case, the knowledge of $\vec{x}^*$ allows to obtain the expectation value $\langle X\rangle^*$ and standard deviation $\sigma^*[X]$ of any weighted topological property $X$ analytically across the ensemble of weighted graphs with, on average, the same strength sequence $\vec{s}(\mathbf{W}^*)$ as the real network $\mathbf{W}^*$. Again, the time required to obtain $\langle X\rangle^*$ is as short as that required to measure the empirical value $X(\mathbf{W}^*)$, as $\langle X\rangle^*$ can be obtained by replacing $w_{ij}^*$ with the expectation value 
\begin{equation}
\langle w_{ij}\rangle^*=\frac{x^*_i x^*_j}{1-x^*_i x^*_j}
\label{eq_wcorrect}
\end{equation}
into the definition of $X(\mathbf{W}^*)$. Equation (\ref{eq_wcorrect}) corrects the naive expectation (\ref{eq_clw}).

In order to apply our method, we need to choose the weighted topological properties to investigate. 
Generalizing binary properties to weighted graphs is arbitrary, as no unique choice exist \cite{vespy_weighted,newman_weighted,myensemble,weightedclustering}.
To better highlight the generality of our approach, here we follow ref.\cite{myensemble} since it introduces a way to always systematically define a weighted counterpart $\tilde{X}$ for every binary property $X$. The idea is to define $\tilde{X}$ as an average of $X$ over the ensemble of binary graphs generated by a convenient connection probability $p_{ij}=f(w_{ij})\in [0,1]$ which is a function of the observed weights $\{w_{ij}\}$. The functional form of $p_{ij}$ can in principle be chosen depending on the empirical properties one wants to detect; however, our purpose here is using our method to compare the empirical properties with the expected ones, rather than comparing alternative definitions of the empirical properties themselves. Therefore we make the simplest choice and, given a real weighted network $\mathbf{W}^*$, we set $p_{ij}\equiv w^*_{ij}/W^*$ where $W^*\equiv\sum_{i<j}w^*_{ij}=\sum_i s_i(\mathbf{W}^*)/2$ is the total weight. 
This choice yields the following definition for the weighted degree \cite{myensemble}:
\begin{equation}
\tilde{k}_i(\mathbf{W}^*)=\frac{\sum_{j\ne i}w^*_{ij}}{W^*}=\frac{s_i(\mathbf{W}^*)}{W^*}
\label{eq_kw}
\end{equation}
which is simply proportional to the strength.
Similarly, the weighted ANND and clustering are defined as the counterparts of eqs.(\ref{eq_knn}) and (\ref{eq_c}) \cite{myensemble}:
\begin{equation}
\tilde{k}^{nn}_i(\mathbf{W}^*)\equiv
\frac{\sum_{j\ne i}\sum_{k\ne j} w^*_{ij}w^*_{jk}}
{W^*\sum_{j\ne i} w^*_{ij}}
\label{eq_knnw}
\end{equation}
\begin{equation}
\tilde{c}_i(\mathbf{W}^*)\equiv
\frac{\sum_{j\ne i}\sum_{k\ne i,j}w^*_{ij}w^*_{jk}w^*_{ki}}
{W^*\sum_{j\ne i}\sum_{k\ne i,j}w^*_{ij}w^*_{ki}}
\label{eq_cw}
\end{equation}
In analogy with the binary case,  $\tilde{k}^{nn}_i$ and $\tilde{c}_i$ can be plotted against $\tilde{k}_i$ (or equivalently $s_i$) in order to investigate the correlation structure of the weighted network.

\begin{figure}
\begin{center}
\includegraphics[width=.48\textwidth]{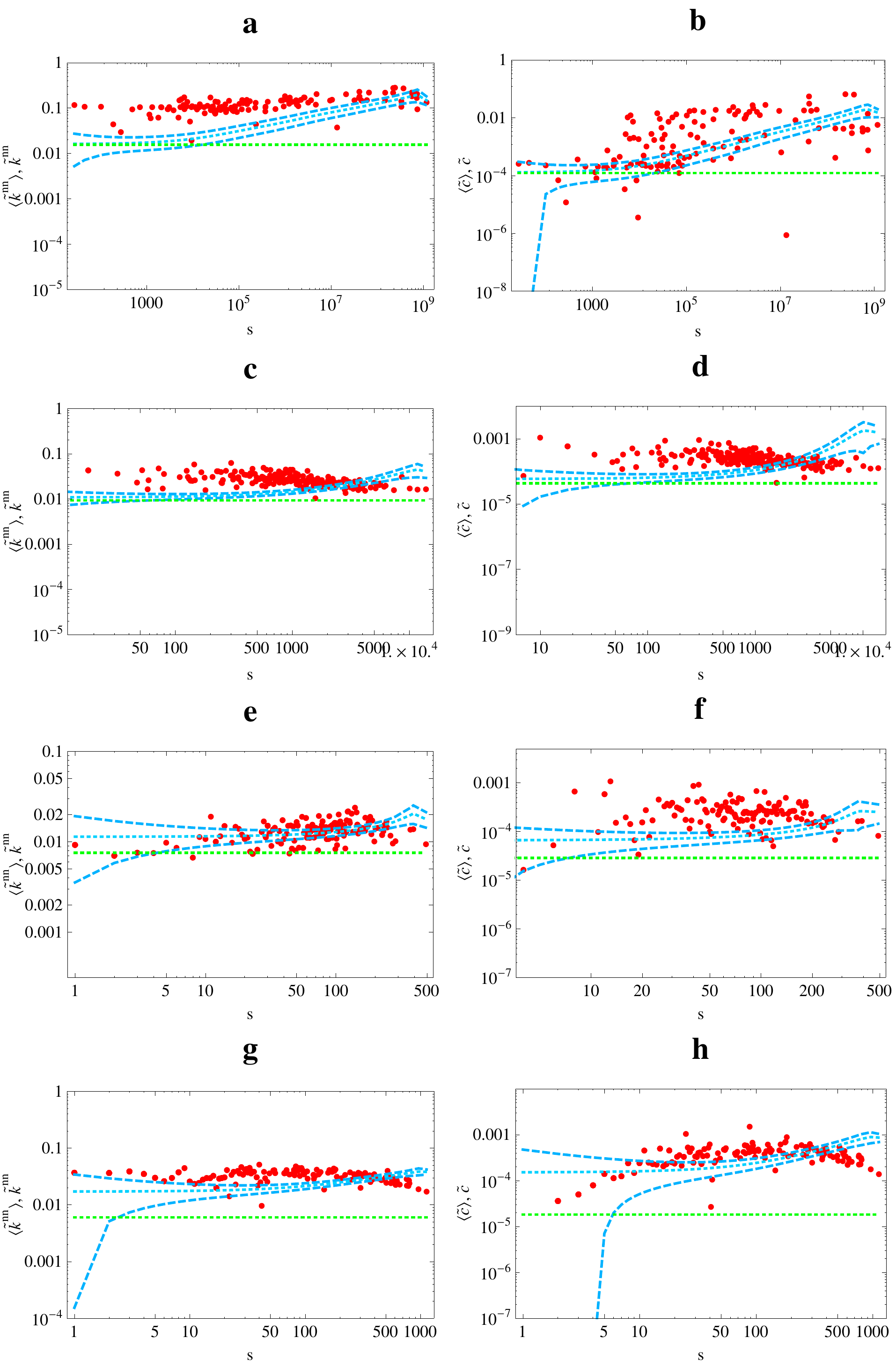}
\end{center}
\caption{Application of our method to weighted undirected networks. Red points are the empirical data and the blue dashed curves are the exact expectations obtained using our method ($\pm$ one standard deviation). Green dashed curves are the flat expectations under the \textit{weighted random graph model}, WRG \cite{myWRG}.
The panels report $\tilde{k}^{nn}_i$ versus $s_i$ (left) and $\tilde{c}_i$ versus $s_i$ (right) for: a) and b) the Florida Bay food web (N=128) \cite{pajek}, c) and d) the Italian interbank network (N=215) \cite{interbank}, e) and f) the \emph{C. elegans} neural network (N=265) \cite{neural}, g) and h) a  snapshot of the US airport network (N=332) \cite{airportdata}.\label{fig_weighted}}
\end{figure}

In fig.\ref{fig_weighted} we analyze the weighted and undirected (symmetrized) versions of four networks we have already considered in the previous binary study: the the Florida Bay food web, the Italian interbank network, the C. Elegans neural network and the US airport network. We compare the empirical results with the expected trends ($\pm$ one standard deviation) under the WCM obtained using our method.
For simplicity, we only show the results obtained using our method, and omit the time-consuming microcanonical comparison.
Note that, since the strengths are preserved in the WCM, i.e. $\langle s_i\rangle^*=s_i(\mathbf{W}^*)$ $\forall i$, the total weight is preserved as well: $\langle W\rangle^*=W^*$.
We find that the empirical trends are quite scattered and variable: some are weakly increasing (Florida Bay), some are approximately constant (interbank web), others first increase and then decrease (airport network).
These diverse trends must be compared with a null model which, unlike naively expected from eq.(\ref{eq_clw}), is not flat and displays a not easily characterizable increasing behavior. 
A common feature is that, with respect to the null behavior, real weighted networks are more assortative and clustered for low values of the strength, while they are less assortative and clustered for high values for the strength. 
These considerations confirm that, even in the weighted case, the empirical trends are uninformative by themselves, and always require a comparison with a null model. Our method allows to treat the otherwise problematic WCM in a simple way, in straightforward analogy with the binary case. 

Although we do not consider this possibility here explicitly, for weighted networks one could also enforce additional constraints on the degree sequence. This amounts to specifying not only the strength of each vertex, but also its purely topological degree \cite{serrano_weighted,colizza2,manna}. In this case, sampling the randomized ensemble by means of computational algorithms becomes even more difficult. By contrast, our method can still be used efficiently, as the analytical expressions characterizing the corresponding maximum-entropy ensemble have been derived recently \cite{mybosefermi}. Those results easily allow to obtain the equations implied by the ML principle, as well as the expectation values of network properties over the ensemble, in a straightforward fashion.

\begin{figure}[h!]
\begin{center}
\includegraphics[width=.48\textwidth]{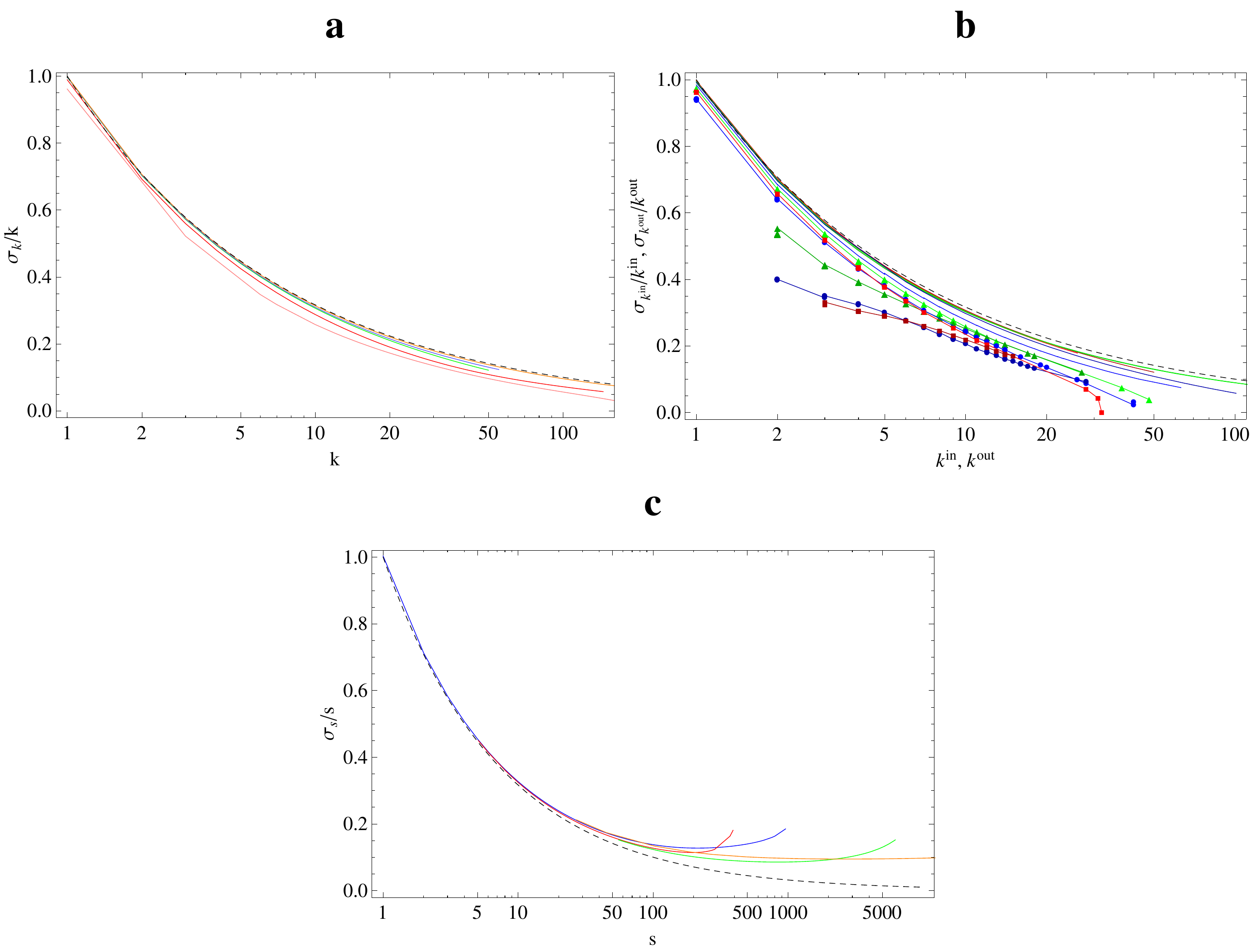}
\end{center}
\caption{The panels report a) the ratios $\sigma^*[k_{i}]/k_{i}$ plotted versus the degrees $k_{i}$ for the binary undirected networks of fig.\ref{fig_undirected}, b) the ratios $\sigma^*[k_{i}^{out}]/k_{i}^{out}$ and $\sigma^*[k_{i}^{in}]/k_{i}^{in}$ plotted versus the degrees $k_{i}^{out}$ and $k_{i}^{in}$, respectively, for the binary directed networks of fig.\ref{fig_directed1} and fig.\ref{fig_foodwebs} and c) the ratios $\sigma^*[s_{i}]/s_{i}$ plotted versus the strengths $s_{i}$ for the weighted undirected networks of fig.\ref{fig_weighted}. The food webs are indicated by means of symbols. The black dashed line is the function $f(x)=x^{-1/2}$ which is expected to well reproduce the coefficients of variation for small values of the constraints.
\label{fig_errk}}
\end{figure}

For completeness, in fig.\ref{fig_errk} we show the ratios of the contraints standard deviations, $\sigma_{\vec{C}}$, to the constraints expected values, $\mu_{\vec{C}}$ (a quantity known in statistics as \textit{coefficient of variation}), plotted versus the expected values. For small values of the constraints $\sigma_{\vec{C}}/\mu_{\vec{C}}\sim (\mu_{\vec{C}})^{-1/2}$ (an approximation valid both for binary and weighted networks); the higher the constraints expected values, the more important becomes a correction factor whose entity (and sign) depends on the particular type of network considered (see Appendix for the details of the calculations): in the food webs (panel b) the presence of in-degree hubs implies the correction to be important even for small out-degree vertices.

\section{Discussion}
Our method make use of the correct expressions (\ref{eq_acorrect}) and (\ref{eq_wcorrect}) for the connection probability and expected weight respectively, in place of the incorrect naive expressions (\ref{eq_cl}) and (\ref{eq_clw}). 
While the latter depend only on the properties ($k_i$ or $s_i$) of the end-point vertices $i$ and $j$, the former depend on the entire degree or strength sequence through eqs.(\ref{eq_k}) and (\ref{eq_s}).
We have shown that this has a dramatic effect on the properties of the randomized ensemble. In particular, we have found that enforcing the same set of constraints in different networks can yield very different trends for the randomized properties, whose behavior is therefore highly unpredictable \emph{a priori}. The general expectation that randomized higher-order properties (such as $\langle k^{nn}_i\rangle$ and $\langle c_i\rangle$ in unweighted networks or $\langle \tilde{k}_i\rangle$ and $\langle \tilde{c}_i\rangle$ in weighted networks) are independent of the local ones ($k_i$ or $s_i$) turns out to be only a very infrequent possibility among the possible scenarios. Indeed, we have also found increasing and decreasing trends for the randomized quantities, and shown that the particular behavior displayed by the null model highly depends on the particular values of the constraints in the original real-world network.
This makes the comparison with the particular null model even more important than previously expected, and underlines the importance of a tractable description enabled by our analytical method.

The incorrectness of eqs.(\ref{eq_cl}) and (\ref{eq_clw}), as well as of their directed counterparts, has another series of undesired effects, as those expressions have been explicitly used to define important structural quantities involved in network analysis. 
Indeed, even when not explicitly used to randomize a network, null models unavoidably enter into the analytical expressions defining many properties of interest.
For instance, many popular community detection algorithms make use of the concept of \emph{modularity} to evaluate the quality of a partition of the network against a null case \cite{santo}. 
A partition into communities can be represented by the matrix $\{\delta_{ij}\}$, where $\delta_{ij}=1$ if vertices $i$ and $j$ are assigned the same community and $\delta_{ij}=0$ otherwise. For a binary undirected network $\mathbf{A}^*$, the modularity $Q$ of the partition $\{\delta_{ij}\}$ has been defined as
\begin{equation}
Q\equiv \frac{1}{2L^*} \sum_{i\ne j}\delta_{ij}\left(a^*_{ij}-p_{ij}\right)
\label{eq_Q1}
\end{equation}
where $p_{ij}$ is the probability that $i$ and $j$ are connected in a suitable null model, and the most frequent choice is the CM. 
Similarly, for a weighted undirected network $\mathbf{W}^*$ the modularity of the partition $\{\delta_{ij}\}$ is \cite{newman_weighted}
\begin{equation}
Q\equiv \frac{1}{2W^*} \sum_{i\ne j}\delta_{ij}\left(w^*_{ij}-\langle w_{ij}\rangle\right)
\label{eq_Q2}
\end{equation}
where $\langle w_{ij}\rangle$ is the expected weight of the link joining $i$ and $j$ in the WCM. 
Unfortunately, the expressions for $p_{ij}$ and $\langle w_{ij}\rangle$ are always taken to be eqs.(\ref{eq_cl}) and (\ref{eq_clw}) respectively. To the best of our knowledge, no rigorous assessment of the consequences of using these approximations has been provided. Therefore the problems described in the present paper affect any modularity-based community detection problem in an uncontrolled way. 
Our methods provides the previously unavailable exact expressions (\ref{eq_acorrect}) and (\ref{eq_wcorrect}), whose values can be inserted into eqs.(\ref{eq_Q1}) and (\ref{eq_Q2}) to have the correct modularity.
A straightforward analysis of how the correct expressions  change the detected community structure of real networks is an important open point to address in the future.

\section{Conclusions}
We have presented a fast and exact method to obtain analytical results about the grandcanonical ensemble of randomized variants of a particular real-world network that preserve its average local properties. The method works for both weighted and unweighted networks, and for both directed and undirected graphs. In any case, it requires as the input only the strength or degree sequence(s), which represent the sufficient statistics of the problem. Our approach can be extended to enforce different or additional constraints, such as the reciprocity structure in directed networks or the simultaneous specification of strengths and degrees in weighted networks. Notably, our results show that maximally random networks exhibit a diverse range of behaviors which is sensitive to the particular values of the constraints displayed by the real network, making a  case-by-case comparison of the observed properties with the randomized ones necessary.
This diversity of outcomes is in any case not captured by widely used but incorrect expressions for the expected properties. Unfortunately, important network properties such as the modularity completely rely on such expressions, a problem that may have therefore biased previous analyses of community structure in networks.
We believe that our contribution represents a promising step towards the identification of relevant information in real networks.

\appendix
\section{\uppercase{General maximum-likelihood method}\label{app_general}}

Here we describe our maximum-likelihood method in its general formulation. Our approach combines previous analytical results (obtained by one of us \cite{mygrandcanonical,mybosefermi,myWRG} and other authors \cite{newman_origin,newman_expo,ginestra_entropy}) about the  properties of maximum-entropy graph ensembles with previous results (by one of us \cite{mylikelihood}) about the maximum-likelihood estimation of free parameters in such ensembles, and adds to them a new technique to obtain analytical expressions for the expectation value and standard deviation of any topological property of interest across the ensemble.
After describing the method in general terms, we derive the explicit expressions that apply in the particular cases of local constraints (for undirected, directed and weighted networks). We then consider an extension to nonlocal constraints, and finally compare our analytical method with alternative computational techniques. 

\subsection{Maximum-entropy probability distribution\label{app_maxent}}
Our method aims at characterizing analytically the properties of families of randomized variants of a particular real network. 
In more rigorous terms, a family of randomized network variants is a statistical ensemble of graphs where a set of structural constraints has been specified, and the rest of the topology is completely random.
Let us denote by $\mathbf{G}$ a generic network in the ensemble, and by ${\mathbf{G}^*}$ the particular real-world network that we need to randomize. The ensemble will consist of all possible networks $\{\mathbf{G}\}$ of the same type of ${\mathbf{G}^*}$ (binary/weighted, directed/undirected), and will include ${\mathbf{G}^*}$ itself.
For binary (either directed or undirected) networks, each graph $\mathbf{G}$ is completely specified by its adjacency matrix $\mathbf{A}$, i.e. $\mathbf{G}\equiv\mathbf{A}$. Similarly, for weighed (either directed or undirected) networks, each graph $\mathbf{G}$ is completely specified by its weight matrix $\mathbf{W}$, i.e. $\mathbf{G}\equiv\mathbf{W}$. We will keep our discussion completely general and use $\mathbf{G}$ to indicate a graph of any type (directed/undirected, binary/weighted).
Thus $\mathbf{G}$ can always be thought of as a matrix with entries $\{g_{ij}\}$, where $g_{ij}$ represents the (either binary or non-negative) weight of the edge $(i,j)$. Any  topological property $X$ evaluates to $X(\mathbf{G})$ when measured on the particular network $\mathbf{G}$, i.e. it is an (arbitrarily complicated) function of the entries $\{g_{ij}\}$.

Each graph $\mathbf{G}$ in the ensemble has an occurrence probability $P(\mathbf{G})$ whose form is determined by the particular constrains enforced. 
This probability must always be such that
\begin{equation}
\sum_\mathbf{G}P(\mathbf{G})=1
\end{equation}
where the sum runs over all graphs in the ensemble.
The expectation value of a topological property $X$ across the ensemble is the mean value (ensemble average)
\begin{equation}
\langle X\rangle\equiv \sum_\mathbf{G} X(\mathbf{G}) P(\mathbf{G})
\label{eq_X}
\end{equation}
Let us denote the set of constraints $\{C_a\}$ by the vector $\vec{C}$, where each $C_a$ is a topological property that, unlike any other generic property $X$, we need to tune to the particular value displayed by the real network ${\mathbf{G}^*}$. 
Enforcing the constraints \emph{exactly}, i.e. allowing only the graphs $\mathbf{G}$ such that $\vec{C}(\mathbf{G})=\vec{C}({\mathbf{G}^*})$, results in a so-called \emph{microcanonical} ensemble characterized by the uniform probability 
\begin{equation}
P(\mathbf{G})=\left\{\begin{array}{ll}1/\mathcal{N}[\vec{C}({\mathbf{G}^*})]&\textrm{   if   }\vec{C}(\mathbf{G})=\vec{C}({\mathbf{G}^*})\\
0&\textrm{   otherwise}\end{array}\right.
\end{equation}
where $\mathcal{N}[\vec{C}({\mathbf{G}^*})]$ denotes the number of graphs in the ensemble for which the value of each constraint $C_a$ equals the value $C_a({\mathbf{G}^*})$. 
Microcanonical graph ensembles are hard to deal with analytically, and they are most often sampled computationally by generating many randomized networks explicitly, using probabilistic rules that ensure that the constraints are matched exactly. 
Currently, such computational techniques are the only available method to randomize a real network. Unfortunately, the need to sample the ensemble explicitly and generating a large number of randomized graphs makes this approach computationally demanding, time consuming and beyond analytic control.

In order to develop a randomization method which is fast and analytically tractable, we exploit the results in ref.\cite{newman_expo} and consider the alternative possibility to enforce the constraints \emph{on average}, i.e.  by only specifying their expectation values $\langle\vec{C}\rangle$.
The resulting ensemble is a \emph{(grand)canonical} one where each graph $\mathbf{G}$ is assigned a probability $P(\mathbf{G})$ that maximizes the Shannon-Gibbs entropy
\begin{equation}
S\equiv -\sum_\mathbf{G} P(\mathbf{G})\ln P(\mathbf{G})
\label{eq_entropy}
\end{equation}
subject to the constraints $\sum_\mathbf{G}P(\mathbf{G})=1$ and $\langle\vec{C}\rangle=\vec{C}$. 
Maximizing the entropy subject to constraints is widely used in statistical mechanics, and in general for problems with incomplete information \cite{jaynes1,jaynes2}.
The desired maximum-entropy graph probability can be found by introducing a set of Lagrange multipliers $\vec{\theta}=\{\theta_a\}$ enforcing the constraints  $\vec{C}=\{C_a\}$. The resulting conditional (on the value of $\vec{\theta}$) probability reads \cite{newman_expo}
\begin{equation}
P(\mathbf{G}|\vec{\theta})=\frac{e^{-H(\mathbf{G},\vec{\theta})}}{Z(\vec{\theta})}
\label{eq_grandP}
\end{equation}
where $H(\mathbf{G},\vec{\theta})$ is the \emph{graph Hamiltonian} defined as the linear combination
\begin{equation}
H(\mathbf{G},\vec{\theta})\equiv\sum_a \theta_a C_a(\mathbf{G})=\vec{\theta}\cdot\vec{C}(\mathbf{G})
\label{eq_H}
\end{equation}
and the normalizing quantity $Z(\vec{\theta})$ is the \emph{partition function}, defined as
\begin{equation}
Z(\vec{\theta})\equiv\sum_\mathbf{G} e^{-H(\mathbf{G},\vec{\theta})}
\label{eq_partition}
\end{equation}
The above results show that the graph probability $P(\mathbf{G}|\vec{\theta})$ always depends on the value $\vec{\theta}$, which in turn depends on the constraints considered. As a consequence, we can rewrite eq.(\ref{eq_X}) more explicitly as a function of $\vec{\theta}$:
\begin{equation}
\langle X\rangle_{\vec{\theta}}\equiv \sum_\mathbf{G} X(\mathbf{G}) P(\mathbf{G}|\vec{\theta})
\label{eq_X2}
\end{equation}
where $\langle \cdot\rangle_{\vec{\theta}}$ denotes that the ensemble average is evaluated at the particular parameter choice $\vec{\theta}$.
The above expression clarifies that the expectation value of any topological property $X$ depends on the specific enforced constraints through $\vec{\theta}$. Different choices of the constraints imply different values of $\vec{\theta}$, $P(\mathbf{G}|\vec{\theta})$, and $\langle X\rangle_{\vec{\theta}}$.

\subsection{Maximum-likelihood parameter estimation\label{app_maxlike}}
As we mentioned, maximum-entropy graph ensembles generated by eq.(\ref{eq_grandP}) have been used extensively to characterize mathematically networks with specified properties \cite{newman_origin,mygrandcanonical,newman_expo,ginestra_entropy,mybosefermi}. 
However, previous studies did not focus on the randomization of a particular real network (which is our main interest here), but rather on the effects that the specification of certain structural properties has on other aspects of network topology. 
As a consequence, the Lagrange multipliers $\{\theta_a\}$ have been considered as free parameters, generally drawn from carefully chosen probability densities \cite{newman_expo,ginestra_entropy,mybosefermi} that allow analytical results, in terms of which the properties of the network model have been investigated. 
In most cases, the aim has been to explore the topological properties in the thermodynamic limit $N\to\infty$, where $N$ is the number of vertices of the network. 
This means that only generic statistical properties of real networks, such as a power-law degree distribution, were used to generate the ensemble. 
However, this implies that the specific properties of a particular real network (such as deviations of individual vertices from the fitted degree distribution, the intrinsic finiteness of the system, etc.) are ignored and, more importantly, that there is no correspondence between the vertices of the real network and those of the model. 
Thus this approach allows to inspect the properties of maximum-entropy graph ensembles, but does not allow the latter to be considered as null models of \emph{a particular real network}. 
As a consequence, it cannot be used to  detect empirical topological patterns consisting of statistically significant deviations from a null network model.

Here we make one step forward and construct, for a given choice of the constraints, the particular maximum-entropy graph ensemble representing the family of correctly randomized counterparts of a given real network ${\mathbf{G}^*}$. 
Explicitly, we consider a grandcanonical ensemble of graphs with the same number $N$ of vertices as the real network, and for a given choice of the constraints we fit the model defined by eq.(\ref{eq_grandP}) to the empirical network ${\mathbf{G}^*}$.
To this end, we exploit previous results \cite{mylikelihood} showing that maximum-entropy graph ensembles defined by eq.(\ref{eq_grandP}) are a particular class of models for which the maximum-likelihood principle provides an excellent method of parameter estimation, since they are free from problems of bias afflicting other network models.  
In particular, it can be easily shown \cite{mylikelihood} that the  log-likelihood 
\begin{equation}
\mathcal{L}(\vec{\theta})\equiv \ln P({\mathbf{G}^*}|\vec{\theta})=-H({\mathbf{G}^*},\vec{\theta})-\ln Z(\vec{\theta})
\label{eq_likelihood}
\end{equation}
to obtain the real network ${\mathbf{G}^*}$ is maximized by the particular parameter choice $\vec{\theta}^*$ such that the ensemble average $\langle C_a\rangle_{\vec{\theta}^*}$ of each constraint $C_a$ equals the empirical value $C_a({\mathbf{G}^*})$ measured on the real network:
\begin{equation}
\langle\vec{C}\rangle^*\equiv \langle\vec{C}\rangle_{\vec{\theta}^*}=\sum_\mathbf{G}\vec{C}(\mathbf{G})P(\mathbf{G}|\vec{\theta}^*)
=\vec{C}({\mathbf{G}^*})
\label{eq_caverage}
\end{equation}
where we have used $\langle\cdot\rangle^*$ as a shorthand notation to indicate the ensemble average $\langle\cdot\rangle_{\vec{\theta}^*}$ evaluated at the particular value $\vec{\theta}^*$.
The above results means that the maximum likelihood principle indicates, for maximum-entropy graph ensembles, precisely the parameter choice that ensures that the desired constraints are met. This is not true in general: in other network models, tuning the average values of the topological properties of interest to their empirical values requires a parameter choice which in general does not maximize the likelihood to obtain the real network \cite{mylikelihood}, thus introducing a bias in the analysis.

The idea to take the observed constraints $\vec{C}(\mathbf{A}^*)$ as the input and find the `hidden' values $\vec{\theta}^*$ that generate those constraints as the most probable ones was already proposed in ref.\cite{mylikelihood} with the purpose of checking whether $\vec{\theta}^*$ correlates with some external set of empirical non-topological quantities, thus unveiling possible mechanisms shaping the network topology.
Here we make progress, noting that finding the values $\vec{\theta}^*$ represents a preliminary step in order to generate the randomized ensemble we are looking for, and have complete analytic control over it. This is completely independent of whether there are external empirical quantities correlating with $\vec{\theta}^*$.

Note that in eqs.(\ref{eq_X2}) and (\ref{eq_caverage}) the expectation values and the model parameters play inverted roles: while in eq.(\ref{eq_X2}) the expectation values are obtained as a function of the parameters $\vec{\theta}$ which can be varied arbitrarily, in eq.(\ref{eq_caverage}) the observed constraints, which are measured on the particular real network and are therefore given as an input, are used to fix the model parameters to the values $\vec{\theta}^*$.
Interestingly, this opposite line of research has been used quite extensively in traditional social network analysis (where  maximum-entropy ensembles of networks  are widely used under the names of $p^*$, logit or exponential random graph models \cite{HL,WF,tom}) but has not yet been transferred to the randomization problem frequently occurring in complex networks theory.
As we show below, the maximum-likelihood parameter choice is exactly what we need in order to obtain statistically correct expectations over ensembles of randomized variants of any particular real-world network. 
This allows to understand which properties of a real-world network can be simply traced back to the enforced constraints, and which require more complicated explanations.
Another important difference with respect to the main approach followed in social network analysis is that our method allows to analyze weighted networks in exactly the same way as binary graphs, which are instead usually not studied within the $p^*$ framework. 
As a consequence, some of the analytical results we derive and use represent previously unavailable tools to study weighted networks (and maximum-entropy ensembles of them) through a straightforward analogy with binary networks.
Finally, in all the applications we consider it is always possible to find the maximum-likelihood parameter values $\vec{\theta}^*$ exactly even for large networks, without resorting to the approximate techniques traditionally used in social network analysis \cite{tom}.
Therefore our approach extends in many directions the connection between exponential random graphs and maximum-entropy network ensembles, and strengthens considerably the existing relation between social science and network theory.

\subsection{Expectation values of topological properties\label{app_expectation}}
Equation (\ref{eq_caverage}) provides an implicit expression for the value $\vec{\theta}^*$, and solving it is equivalent to maximizing eq.(\ref{eq_likelihood}). The numerical value of the solution $\vec{\theta}^*$ is the key ingredient we are looking for in order to detect topological patterns in the real network ${\mathbf{G}^*}$ analytically, without performing any time-consuming computational randomization. 
Indeed, if we insert the value $\vec{\theta}^*$ into eq.(\ref{eq_X2}) we obtain
\begin{equation}
\langle X\rangle^*\equiv
\langle X\rangle_{\vec{\theta}^*}=
\sum_\mathbf{G}X(\mathbf{G})P(\mathbf{G}|\vec{\theta}^*)
\label{eq_X*}
\end{equation}
which provides the exact expected value of any topological property $X$ across the maximum-entropy graph ensemble where the expected values $\langle \vec{C}\rangle$ of the topological properties $\vec{C}$ chosen as constraints are set equal to the empirical values $\vec{C}({\mathbf{G}^*})$ measured on the real network ${\mathbf{G}^*}$, as ensured by eq.(\ref{eq_caverage}). 
For simplicity, given a real network ${\mathbf{G}^*}$ and a set of constraints $\vec{C}$, we shall sometimes call $\langle X\rangle^*$ the \emph{randomized value} of the topological property $X$.
The comparison of $\langle X\rangle^*$ with the empirical value $X({\mathbf{G}^*})$ allows to assess whether, in the real network $\textbf{{G}}^*$, the topological property $X$ requires additional information besides that provided by the properties $\vec{C}$. 
If $X({\mathbf{G}^*})$ is sufficiently close to $\langle X\rangle^*$ (within a statistical error that we determine in \ref{app_variance}), one can conclude that the enforced constraints $\vec{C}$ fully explain the property $X$.
By contrast, if $X({\mathbf{G}^*})$ is significantly different from $\langle X\rangle^*$, then the properties $\vec{C}$ do not explain the property $X$, which means that the structure of ${\mathbf{G}^*}$ is determined by other factors besides those determining $\vec{C}$.
This allows to assess which topological properties can be traced back to (i.e. explained by) the chosen constraints in any real network, and which can not. 
Trivially, if $X$ is one of the properties among the enforced constraints (i.e. if $X=C_a$ for some $a$), then eq.(\ref{eq_caverage}) implies $X({\mathbf{G}^*})=\langle X\rangle^*$ by construction.

Note that any other parameter choice $\vec{\theta}\ne\vec{\theta}^*$ would not enforce the chosen constraints and would yield an expectation value $\langle X\rangle_{\vec{\theta}}$ different from the desired one, i.e. not corresponding to the correct randomized value $\langle X\rangle^*$ for that particular network and for that particular choice of the constraints.
This clarifies why previous results \cite{newman_origin,mygrandcanonical,newman_expo,ginestra_entropy,mybosefermi} about the properties of maximum-entropy ensembles, that were obtained using $\vec{\theta}$ as a free parameter unrelated to the empirical values $\vec{C}({\mathbf{G}^*})$ and to the real network ${\mathbf{G}^*}$ itself, cannot be used in order to solve the pattern detection problem considered here.
Also note that $\vec{C}({\mathbf{G}^*})$ is the \emph{sufficient statistics} of our problem, which completely determines ${\theta}^*$ through eq.(\ref{eq_caverage}) and consequently any randomized property $\langle X\rangle^*$. The knowledge of the other topological properties of the real network ${\mathbf{G}^*}$ is useless. This means that two real networks $\mathbf{G}_1^*$ and $\mathbf{G}_2^*$ with exactly the same values $\vec{C}(\mathbf{G}_1^*)=\vec{C}(\mathbf{G}_2^*)$ of the constraints generate the same maximum-entropy ensemble, and give rise to the same value of ${\theta}^*$ and $\langle X\rangle^*$, as should be.

Clearly, the possibility to solve eq.(\ref{eq_caverage}) and to obtain the randomized properties through eq.(\ref{eq_X*}) both depend on whether one manages to rewrite the formal expression for $\langle X\rangle_{\vec{\theta}}$ in eq.(\ref{eq_X2}) in a simplified form that avoids the unfeasible actual enumeration of all graphs $\{\mathbf{G}\}$ in the ensemble. 
In practical terms, this means that not all specifications of the constraints $\vec{C}$ allow to solve eq.(\ref{eq_caverage}) and obtain $\vec{\theta}^*$, and not all topological properties $X$ allow to be averaged exactly through eq.(\ref{eq_X*}).
However, as we describe in \ref{app_local}, the first step can always be carried out successfully whenever one considers \emph{local} constraints as the ones of interest for us. Similarly, as we now show in general and then restate more explicitly in each particular case, the expectation value $\langle X\rangle^*$ of any higher-order topological property $X$ of interest can be rewritten, either exactly or approximately, in a way that is only as complicated as measuring $X({\mathbf{G}^*})$ on a single network, rather than on all graphs $\{\mathbf{G}\}$ in the ensemble. This represents a major advantage of our method: the computation of an expectation value across the entire ensemble of graphs is only as time-consuming as the computation of the corresponding observed value on the empirical network  ${\mathbf{G}^*}$. Thus, if the observed value can be computed in reasonable time, the same is true for the expectation value.
To see this, we write down an approximated expression for $\langle X\rangle^*$ as a Taylor expansion. Note that any property $X(\mathbf{G})$ depends in general on all the entries $\{g_{ij}\}$ of the matrix $\mathbf{G}$, which are the fundamental degrees of freedom of the problem. 
The ensemble average of $g_{ij}$ reads
\begin{equation}
\langle g_{ij}\rangle=\sum_\mathbf{G}g_{ij}P(\mathbf{G}|\vec{\theta})
\label{eq_gmean_no*}
\end{equation}
If we define the gradient matrix of any topological property $X(\mathbf{G})$ as
\begin{equation}
\mathbf{\nabla}X(\mathbf{G})\equiv\left(\begin{array}{ccc}\frac{\partial X(\mathbf{G})}{\partial g_{11}}&\dots&\frac{\partial X(\mathbf{G})}{\partial g_{1N}}\\
\vdots& &\vdots\\
\frac{\partial X(\mathbf{G})}{\partial g_{N1}}&\dots&\frac{\partial X(\mathbf{G})}{\partial g_{NN}}\end{array}\right)
\end{equation}
and if we denote by $\langle\mathbf{G}\rangle$ the matrix whose entries $\langle\mathbf{G}\rangle_{ij}$ are the expectation values  $\langle g_{ij}\rangle$, it is possible to expand $\langle X\rangle$ around $\langle\mathbf{G}\rangle$ and write the multidimensional first-order Taylor expansion
\begin{eqnarray}
X(\mathbf{G})&=&X(\langle\mathbf{G}\rangle)
+\sum_{i,j}(g_{ij}-\langle g_{ij}\rangle)\left(\frac{\partial X}{\partial g_{ij}}\right)_{\mathbf{G}=\langle\mathbf{G}\rangle}
+\dots\nonumber\\
&=&X(\langle\mathbf{G}\rangle)
+(\mathbf{G}-\langle \mathbf{G}\rangle)*\mathbf{\nabla}X(\langle \mathbf{G}\rangle)+\dots
\label{eq_taylor}
\end{eqnarray}
In the above expression, $(\cdot)_{\mathbf{G}=\langle\mathbf{G}\rangle}$ means that we are evaluating the quantity in brackets by replacing each $g_{ij}$ with $\langle g_{ij}\rangle$, and 
\begin{equation}
\mathbf{A}*\mathbf{B}\equiv\sum_{i,j}a_{ij}b_{ij}
\label{eq_matrixprod}
\end{equation}
denotes the scalar product of two matrices $\mathbf{A}$ and $\mathbf{B}$, and the double sum runs over all $N(N-1)$ ordered pairs of vertices (with $i\ne j$). Note that for an undirected network, where $g_{ij}=g_{ji}$ by construction, half of the terms in the sum in eq.(\ref{eq_taylor}) will be equal to zero, since one has either $\partial X/\partial g_{ji}=0$ or $\partial X/\partial g_{ij}=0$, depending on whether $g_{ij}$ or $g_{ji}$ appears in the formal definition of $X$.
With the above approximation, the expectation value of $X$ reads
\begin{equation}
\langle X\rangle =X(\langle\mathbf{G}\rangle)
+\dots
\end{equation}
since the first-order terms vanish. The above formula shows that, if one evaluates $\langle X\rangle$ by simply replacing $\mathbf{G}$ with $\langle \mathbf{G}\rangle$ into $X(\mathbf{G})$ (\emph{linear approximation}), the difference with respect to the exact expectation value is only in the second- and higher-order terms. This is true for any value of $\vec{\theta}$, on which all expectation values depend. As already explained, our method consists in choosing the particular value $\vec{\theta}^*$ solving eq.(\ref{eq_caverage}), which yields an expectation value 
\begin{equation}
\langle X\rangle^* =X(\langle\mathbf{G}\rangle^*)
+\dots
\label{eq_expXgeneral}
\end{equation}
Among all possible parameter values $\vec{\theta}$, the choice of $\vec{\theta}^*$ ensures that the deviation of the approximate value $X(\langle\mathbf{G}\rangle)$ from the exact one $X(\mathbf{G})$ in eq.(\ref{eq_taylor}) is minimal, since $\langle \mathbf{G}\rangle^*$ is as close as possible to $\mathbf{G}$, if the chosen constraints $\vec{C}$ are chosen as a reference to measure the difference between $\langle \mathbf{G}\rangle$ and $\mathbf{G}$.
In particular, when $X$ coincides with one of the enforced constraints $C_a$, eq.(\ref{eq_expXgeneral}) becomes an exact expression, as we mentioned.
Moreover, as we show later in \ref{app_local}, most topological properties of interest in our analysis are either multilinear functions of statistically independent matrix elements $\{g_{ij}\}$ or ratios of multilinear functions. In the former case, the expectation value $\langle X\rangle^*$ is exactly $X(\langle\mathbf{G}\rangle^*)$. In the latter case, the numerator and denominator will be separately evaluated exactly, and the approximation will only affect the ratio. In general, ratios of averages can be very different from averages of ratios. However we confirmed (see figs.\ref{fig_undirected} and \ref{fig_directed1}) that our estimates for the ratios are in very good accordance with what is obtained in the microcanonical case using the LRA, where averages of ratios are evaluated exactly.
Moreover, recall that we are interested in determining an interval of statistically significant values around $\langle X\rangle^*$, rather than $\langle X\rangle^*$ alone. Our results (figs.\ref{fig_undirected} and \ref{fig_directed1}) also show that the difference between the microcanonical and (approximate) grandcanonical value of $\langle X\rangle^*$ is typically much smaller than the standard deviation of $X$ (that we obtain below), so using eq.(\ref{eq_expXgeneral}) is in any case a very good  way to proceed. 

The above discussion clarifies that a good approximation to the randomized value $\langle X\rangle^*$ of any topological property of interest is given by simply replacing each $g_{ij}$ with $\langle g_{ij}\rangle^*$ in the definition of the property $X(\mathbf{G})$, in the same way as the empirical value $X({\mathbf{G}^*})$ is obtained by replacing each $g_{ij}$ with the observed entry $g_{ij}^*$ of ${\mathbf{G}^*}$ in the definition of $X(\mathbf{G})$. 
This means that the empirical value $X({\mathbf{G}^*})$ (if the full adjacency matrix is used, see main text) and the approximate randomized value $X(\langle\mathbf{G}\rangle^*)$ require exactly the same computational time, which makes our method faster than any other available alternative approach (and in general as fast as possible). 
Clearly, in order to evaluate eq.(\ref{eq_expXgeneral}) the complete knowledge of the values 
\begin{equation}
\langle g_{ij}\rangle^*=\sum_{\mathbf{G}}g_{ij}P(\mathbf{G}|\vec{\theta}^*)
\label{eq_gmean}
\end{equation}
is required. While for generic choices of $\vec{C}$ it may be impossible to obtain the formal expression for $\langle g_{ij}\rangle_{\vec{\theta}}$ and/or the particular parameter value $\vec{\theta}^*$, in \ref{app_local} we show that local constraints always allow to obtain $\langle g_{ij}\rangle^*$ exactly. This makes the problem analytically solvable, and implies that our method becomes very simple in all the applications of interest.

\subsection{Variances of topological properties\label{app_variance}}
As we mentioned, another important advantage of our method is the possibility to obtain, besides the expectation value, the analytical expression for the standard deviation of any topological property of interest. This provides a statistical error allowing to detect significant deviations of any empirically observed topological quantity $X({\mathbf{G}^*})$ from its randomized value $\langle X\rangle^*$.
To this end, we employ the fundamental expression relating the variance of a function of many random variables to the variances of the latter, whose most popular consequence is the general formula for the propagation of errors in experimental measurements. 
In our notation, the variance of a topological property $X$ across the ensemble is defined as
\begin{equation}
\sigma^2[X]\equiv
\langle X^2\rangle-\langle X\rangle^2=\langle(X-\langle X\rangle)^2\rangle
\label{eq_variancedef}
\end{equation}
(which depends on $\vec{\theta}$). Using the linear approximation in eq.(\ref{eq_taylor}) we can write
\begin{eqnarray}
\sigma^2[X]&=&\langle[X(\mathbf{G})-X(\langle\mathbf{G}\rangle)]^2\rangle\\
&=&\sum_{i,j}\sum_{t,s}\sigma[g_{ij},g_{ts}]\left(\frac{\partial X}{\partial g_{ij}}\frac{\partial X}{\partial g_{ts}}\right)_{\mathbf{G}=\langle\mathbf{G}\rangle}
+\dots\nonumber
\label{eq_generalpropagation}
\end{eqnarray}
where 
\begin{eqnarray}
\sigma[g_{ij},g_{ts}]&\equiv&\langle(g_{ij}-\langle g_{ij}\rangle)(g_{ts}-\langle g_{ts}\rangle)\rangle\nonumber\\
&=&\langle g_{ij}g_{ts}\rangle-\langle g_{ij}\rangle\langle g_{ts}\rangle
\end{eqnarray}
is the covariance of $g_{ij}$ and $g_{ts}$, and 
\begin{equation}
\langle g_{ij}g_{ts}\rangle=\sum_\mathbf{G}g_{ij}g_{ts}P(\mathbf{G}|\vec{\theta})
\label{eq_ggmean_no*}
\end{equation}
For the `diagonal' terms given by $i=t$ and $j=s$, the covariance $\sigma[g_{ij},g_{ts}]$ equals the variance
\begin{equation}
\sigma^2[g_{ij}]\equiv\langle g_{ij}^2\rangle-\langle g_{ij}\rangle^2=\sigma[g_{ij},g_{ij}]
\label{eq_variancegij}
\end{equation}
(again, both $\sigma[g_{ij},g_{ts}]$ and $\sigma^2[g_{ij}]$ depend on $\vec{\theta}$). 
In a different context where $X$ is a function of many experimental quantities $\{g_{ij}\}$, eq.(\ref{eq_generalpropagation}) provides the general formula for the propagation of errors (from $\{g_{ij}\}$ to $X$), if the measured value of $g_{ij}$ is used as the best estimate for $\langle g_{ij}\rangle$, and if its experimental error is used in place of $\sigma[g_{ij}]$.
Here, we do not need approximate estimates for $\langle g_{ij}\rangle$ and $\sigma[g_{ij}]$, since both quantities can be completely specified: even if there is always a single observation, i.e. the real network ${\mathbf{G}^*}$, the latter generates the entire ensemble of graphs which is described by the probability $P(\mathbf{G}|\vec{\theta}^*)$, as we discussed in detail in \ref{app_maxlike}. 

As for the expectation value $\langle X\rangle^*$, our approach proceeds by evaluating the standard deviation $\sigma[X]^*$ at the particular parameter value $\vec{\theta}^*$ solving eq.(\ref{eq_caverage}):
\begin{equation}
\sigma^*[X]=\sqrt{
\sum_{i,j}\sum_{t,s}\sigma^*[g_{ij},g_{ts}]\left(\frac{\partial X}{\partial g_{ij}}\frac{\partial X}{\partial g_{ts}}\right)_{\mathbf{G}=\langle\mathbf{G}\rangle^*}
+\dots}
\label{eq_starpropagation}
\end{equation}
where 
\begin{equation}
\sigma^*[g_{ij},g_{ts}]=\langle g_{ij}g_{ts}\rangle^*-\langle g_{ij}\rangle^*\langle g_{ts}\rangle^*\mbox{.}
\label{eq_ggstar}
\end{equation}
Note that, as for the expected values, the above standard deviation makes use of the linear approximation and is therefore not exact. However, when we measured also the microcanonical standard deviations, we found an excellent agreement with our grandcanonical ones  (see fig. \ref{fig_directed1}a-b), showing that the errors on the estimates of our standard deviations are small.

Equations (\ref{eq_starpropagation}) and (\ref{eq_ggstar}) show that the values
\begin{equation}
\langle g_{ij}g_{ts}\rangle^*=\sum_{\mathbf{G}}g_{ij}g_{ts}P(\mathbf{G}|\vec{\theta}^*)
\label{eq_ggmean}
\end{equation} 
are the fundamental quantities, besides the averages $\langle g_{ij}\rangle^*$ given by eq.(\ref{eq_gmean}), required in order to obtain the standard deviation $\sigma^*[X]$ of any topological property $X$.
For generic choices of the constraints $\vec{C}$, obtaining the value of $\langle g_{ij}g_{ts}\rangle^*$ can be very complicated or even impossible, as we already discussed for $\langle g_{ij}\rangle^*$. However, as we will show, local constraints always allow to evaluate analytically all the  covariances, and hence the standard deviation $\sigma^*[X]$ of any property $X$.

Equation (\ref{eq_starpropagation}) is the key expression providing the statistical error associated with $\langle X\rangle^*$. For any topological quantity $X$, our method allows to assess whether the empirical value $X({\mathbf{G}^*})$ is consistent with the randomized value $\langle X\rangle^*$ within $z$ standard deviations (where $z$ is a conveniently chosen threshold value), i.e. whether 
\begin{equation}
|X({\mathbf{G}^*})-\langle X\rangle^*|\leq z \sigma^*[X]
\end{equation}
As long as the above inequality holds, it is legitimate to say that the particular property $X$ evidences no significant deviation of the real network ${\mathbf{G}^*}$ from a null model where the constraints $\vec{C}$ are specified. 
This means that the observed value $X({\mathbf{G}^*})$ requires no explanation besides those accounting for the observed values $\vec{C}({\mathbf{G}^*})$ of the constraints. By contrast, if the above inequality is violated, then one has a signature that the observed network ${\mathbf{G}^*}$ is not completely a result of the specification of the constraints $\vec{C}$. 
Additional mechanisms, besides those determining the values of the constraints, are at work. In other words, higher-order patterns which cannot be traced to low-level constraints are present, and our method is able to detect them.
In practice, in order to discriminate between the two possibilities, it is useful to compute the two values 
\begin{equation}
\langle X\rangle^*\pm z\sigma^*[X]
\label{eq_pm}
\end{equation}
which delimit the region within which an observed value $X({\mathbf{G}^*})$ would imply the acceptance of the null model from the one where an observed value $X({\mathbf{G}^*})$ would imply the rejection of the null model.
As an alternative, rather than fixing a threshold value for $z$, one can directly compute the number of standard deviations by which the expected and the empirical value of $X$ differ, i.e. the \emph{z-score}
\begin{equation}
z[X]\equiv\frac{X({\mathbf{G}^*})-\langle X\rangle^*}{\sigma^*[X]}
\label{eq_zscore}
\end{equation}
Large positive (negative) values of $z[X]$ indicate that $X({\mathbf{G}^*})$ is substantially larger (smaller) than expected, while small values signal no significant deviation from the null model (note however, as mentioned in the main text, that $z$-scores are easily interpretable only for normally distributed properties).

This concludes the description of our method in its general form.
In what follows, we consider the particular case of interest for the present analysis, i.e. when the constraints $\vec{C}$ are (either binary or weighted) local topological properties, or when they are nonlocal but simple enough to preserve the analytical character of the method.

\section{\uppercase{Local constraints}\label{app_local}}
The most important case is when the constraints $\vec{C}$ are \emph{local} (or first-order) topological properties, i.e. properties determined by moving only one step away from a vertex, thus reaching only its first neighbours. In binary undirected networks the fundamental local property is the degree $k_i=\sum_{j\ne i}a_{ij}$, while in weighted undirected networks the corresponding quantity is the strength $s_i=\sum_{j\ne i}w_{ij}$. In directed networks, a pair of inward and outward variants of the same quantities (i.e. the in-degree $k_i^{in}$ and out-degree $k_i^{out}$, or the in-strength $s_i^{in}$ and out-strength $s_i^{out}$) characterizes the local properties of each vertex. 
Choosing local constraints is the natural option when one is interested in understanding the effects that the specification of low-order properties, involving only direct interactions, has on higher-order properties involving longer chains of interactions. 
In what follows, we therefore discuss our method in detail in the particular case of local constraints. We will consider both binary and weighted networks, and both undirected and directed links.
Importantly, we will show that in all these cases the graph probability $P(\mathbf{G}|\vec{\theta})$ factorizes as
\begin{equation}
P(\mathbf{G}|\vec{\theta})=\prod_{i<j}D_{ij}(g_{ij},g_{ji}|\vec{\theta})
\label{eq_factorize}
\end{equation}
where the product runs over all unordered pairs of vertices $(i,j)$ (with $i<j$) and $D_{ij}(g,g'|\vec{\theta})$ is the dyadic probability that the pair $(g_{ij},g_{ji})$ takes the particular value $(g,g')$, i.e. the joint probability that $g_{ij}=g$ and simultaneously $g_{ji}=g'$. Clearly,
\begin{equation}
D_{ij}(g,g'|\vec{\theta})=D_{ji}(g',g|\vec{\theta})
\end{equation}
Note that $D_{ij}(g,g'|\vec{\theta})$ is normalized such that
\begin{equation}
\sum_{g,g'}D_{ij}(g,g'|\vec{\theta})=1
\label{eq_Dnorm}
\end{equation}
where $g$ and $g'$ run over all the allowed values for $g_{ij}$ and $g_{ji}$ ($g=0,1$ for binary networks, while $g=0,1\dots+\infty$ for weighted networks; the same for $g'$). 
The marginal probability that $g_{ij}$ takes the particular value $g$, independently of the value of $g_{ji}$, is
\begin{equation}
P_{ij}(g|\vec{\theta})=\sum_{g'}D_{ij}(g,g'|\vec{\theta})
\end{equation}
and, consistently with eq.(\ref{eq_Dnorm}), is normalized such that
\begin{equation}
\sum_{g}P_{ij}(g|\vec{\theta})=1\mbox{.}
\label{eq_Pnorm}
\end{equation}
Note that for undirected networks, where $g_{ij}=g_{ji}$ by construction, we have 
\begin{equation}
D_{ij}(g,g'|\vec{\theta})=\delta_{g,g'}P_{ij}(g|\vec{\theta})
\end{equation}
where $\delta_{g,g'}=1$ if $g= g'$ and $\delta_{g,g'}=0$ if $g\ne g'$.

The factorization of $P(\mathbf{G}|\vec{\theta})$ according to eq.(\ref{eq_factorize}) implies that eq.(\ref{eq_gmean_no*}) can be rewritten as 
\begin{equation}
\langle g_{ij}\rangle=\sum_{g}gP_{ij}(g|\vec{\theta})
\label{eq_gmean_no*2}
\end{equation}
which can always be obtained analytically. 
Using the latter, eq.(\ref{eq_caverage}) can be simply rewritten exactly as
\begin{equation}
\vec{C}(\langle\mathbf{G}\rangle^*)=\vec{C}({\mathbf{G}^*})
\label{eq_caveragelocal}
\end{equation}
which allows the maximum-likelihood parameter values $\vec{\theta}^*$ appearing in $\langle\mathbf{G}\rangle^*$  to be easily calculated numerically. Alternatively (e.g. depending on the software used) one can calculate $\vec{\theta}^*$ by directly maximizing the log-likelihood defined in eq.(\ref{eq_likelihood}), which in this case takes the simpler form
\begin{equation}
\mathcal{L}(\vec{\theta})\equiv \ln P({\mathbf{G}^*}|\vec{\theta})=\sum_{i<j}\ln D_{ij}(g^*_{ij},g^*_{ji}|\vec{\theta})\mbox{.}
\label{eq_likelihoodlocal}
\end{equation}
(we always adopted the maximization of the log-likelihood).
In both cases, even for very large networks this preliminary parameter estimation takes a negligible time with respect to the calculation of any nontrivial topological property. 
This implies that eq.(\ref{eq_gmean_no*2}) can always be evaluated exactly at the particular parameter choice $\vec{\theta}^*$, providing the correct value $\langle g_{ij}\rangle^*$ in terms of which the ensemble average $\langle X\rangle^*$ of any topological property $X$ can be obtained analytically through eq.(\ref{eq_expXgeneral}).
Thus, as we discussed, the time required to obtain $\langle X\rangle^*$ (which formally is an average over all possible graphs in the ensemble) is just the same as that required in order to measure $X({\mathbf{G}^*})$ on the real network ${\mathbf{G}^*}$. 
This makes our method incredibly faster than other randomization procedures that require the actual computational generation of many randomized variants (necessarily sampling only a part of the ensemble) of the real network, on each of which $X$ must be computed explicitly before performing a final average approximating $\langle X\rangle^*$.

The standard deviation $\sigma^*[X]$ of any property $X$ can be evaluated very easily as well. Equation eq.(\ref{eq_factorize}) implies that if $(i,j)$ and $(t,s)$ are two distinct pairs of vertices then
\begin{eqnarray}
&\langle g_{ij}g_{ts}\rangle=\langle g_{ij}\rangle\langle g_{ts}\rangle&\\ &\sigma[g_{ij},g_{ts}]=0&
\label{eq_independence}
\end{eqnarray}
By contrast, if $i=t$ and $j=s$ then 
\begin{eqnarray}
&\langle g_{ij}g_{ij}\rangle=\langle g^2_{ij}\rangle=\sum_{g}g^2 P_{ij}(g|\vec{\theta})&\\
&\sigma[g_{ij},g_{ij}]=\langle g_{ij}^2\rangle-\langle g_{ij}\rangle^2=\sigma^2[g_{ij}]&
\label{eq_independence2}
\end{eqnarray}
Finally, if $i=s$ and $j=t$ we have
\begin{eqnarray}
&\langle g_{ij}g_{ji}\rangle=\sum_{g,g'}g g' D_{ij}(g,g'|\vec{\theta})&
\\
&\sigma[g_{ij},g_{ji}]=\langle g_{ij}g_{ji}\rangle-\langle g_{ij}\rangle\langle g_{ji}\rangle&
\label{eq_independence3}
\end{eqnarray}
Again, all the above quantities can be obtained analytically and evaluated exactly at the particular value $\vec{\theta}^*$ solving eq.(\ref{eq_caveragelocal}). 
As a consequence, if eqs.(\ref{eq_independence}), (\ref{eq_independence2}) and (\ref{eq_independence3}) are inserted into eq.(\ref{eq_generalpropagation}), we find that the expression for the variance $\sigma^2[X]$ of any topological property $X$ reduces from eq.(\ref{eq_starpropagation}) to the simpler formula

\begin{eqnarray}
\label{eq_independentpropagation}
(\sigma^*[X])^2&=&
\sum_{i,j}\left[\left(\sigma^*[g_{ij}]\frac{\partial X}{\partial g_{ij}}\right)_{\mathbf{G}=\langle\mathbf{G}\rangle^*}^2\right.\\
&+&\left.\sigma^*[g_{ij},g_{ji}]\left(\frac{\partial X}{\partial g_{ij}}\frac{\partial X}{\partial g_{ji}}\right)_{\mathbf{G}=\langle\mathbf{G}\rangle^*}\right]
+\dots\nonumber
\end{eqnarray}

involving only a single sum over pairs of vertices.
In the above expression, we have kept our convention to let the sum run always over all possible ordered pairs of vertices, thus considering the pairs $(i,j)$ and $(j,i)$ as distinct terms in the summation. 
For ensembles of directed networks, $g_{ij}$ and $g_{ji}$ are different random variables which may or may not be dependent on each other (depending on the enforced constraints, as we show in detail below). Equation (\ref{eq_independentpropagation}) takes care of both possibilities by including the covariance $\sigma^*[g_{ij},g_{ji}]$. For ensembles of undirected networks, $g_{ij}$ and $g_{ji}$ are actually the same random variable and are thus perfectly correlated, which means $\sqrt{\sigma^*[g_{ij},g_{ji}]}=\sqrt{\sigma^*[g_{ij},g_{ij}]}=\sigma^*[g_{ij}]$. 
Again, eq. (\ref{eq_independentpropagation}) takes care of this by compensating the summation over a doubled number of terms with the presence of the covariances which exactly restore the correct expression. 
In such a way, one does not have to care whether the network is undirected when using eq.(\ref{eq_independentpropagation}), which therefore applies without modifications to all the cases we will consider below. Different cases only differ by the specific expression of $\sigma^*[g_{ij},g_{ji}]$. This is very convenient when implementing the formula computationally.
Another desirable consequence of formally treating $g_{ij}$ and $g_{ji}$ as different variables even in undirected networks is that in eq.(\ref{eq_independentpropagation}) the derivative $\partial X/\partial g_{ts}$ of any function $X(\mathbf{G})$ of (a subset of) the entries $\{g_{ij}\}$ can always be computed by repeatedly applying the elementary rule

\begin{equation}
\frac{\partial g_{ij}}{\partial g_{ts}}=\delta_{it}\delta_{js}
\label{eq_deltarule}
\end{equation}
(where $\delta_{ij}=1$ if $i=j$ and $\delta_{ij}=0$ if $i\ne j$) for both directed and undirected graphs.\\

Summarizing the results discussed so far, we showed that for local constraints our method allows $\langle g_{ij}\rangle^*$, $\langle g_{ij}^2\rangle^*$ and $\langle g_{ij}g_{ji}\rangle^*$ to be computed exactly, and to use them in order to obtain the expected randomized value $\langle X\rangle^*$ and standard deviation $\sigma^*[X]$ of any topological property $X$ through eqs.(\ref{eq_expXgeneral}) and (\ref{eq_independentpropagation}) respectively. Unlike alternative computational methods, our approach is completely analytical and allows to evaluate the randomized value $\langle X\rangle^*$ in just the same time as that required to measure $X$ on the original real network $\mathbf{G}^*$, plus a negligible preliminary time required to find the parameter values $\vec{\theta}^*$ numerically through eq.(\ref{eq_caveragelocal}).
The simple steps through which our method proceeds in the case of local constraints can be summarized as follows:
\begin{enumerate}
\item choose the desired representation for the real network $\mathbf{G}^*$ (directed/undirected, binary/weighted) and the corresponding grandcanonical ensemble of graphs $\{\mathbf{G}\}$;
\item
specify the local constraints $\vec{C}(\mathbf{G})$ and use them to write the Hamiltonian $H(\mathbf{G},\vec{\theta})=\vec{\theta}\cdot\vec{C}(\mathbf{G})$ and the probability 
$P(\mathbf{G}|\vec{\theta})=e^{-H(\mathbf{G},\vec{\theta})}/Z(\vec{\theta})$ according to eqs.(\ref{eq_grandP})-(\ref{eq_partition});
\item rewrite the graph probability analytically in the factorized form $P(\mathbf{G}|\vec{\theta})=\prod_{i<j}D_{ij}(g_{ij},g_{ji}|\vec{\theta})$ according to eq.(\ref{eq_factorize});
\item use $D_{ij}(g,g'|\vec{\theta})$ to determine the basic quantities $\langle g_{ij}\rangle$, $\langle g_{ij}^2\rangle$ and $\langle g_{ij}g_{ji}\rangle$ according to eqs.(\ref{eq_gmean_no*2}), (\ref{eq_independence2}) and (\ref{eq_independence3}) respectively;
\item numerically determine the maximum-likelihood parameters $\vec{\theta}^*$ by solving eq.(\ref{eq_caveragelocal}) or alternatively maximizing eq.(\ref{eq_likelihoodlocal});
\item use $\vec{\theta}^*$ to compute the ensemble average $\langle X\rangle^*$ and standard deviation $\sigma^*[X]$ of any desired topological property $X$, according to eqs.(\ref{eq_expXgeneral}) and (\ref{eq_independentpropagation});
\item assess whether the empirical value $X(\mathbf{G}^*)$ is consistent with the randomized one $\langle X\rangle^*$ using either the interval in eq.(\ref{eq_pm}) or the $z$-score in eq.(\ref{eq_zscore}). 
\end{enumerate}
For completeness, in the above list we have included all the logical steps involving also the initial derivation of the required analytical expressions. However, since those expressions have already been derived in the literature for all the constraints we will consider in what follows, in practice our method reduces to a straightforward application of the last three steps. 
For clarity, in what follows we illustrate the method explicitly for a range of useful specific cases, i.e. for various choices of the constraints $\vec{C}$ and of the topological properties $X$. We will also highlight in more detail the advantages with respect to alternative methods.

\subsection{Undirected configuration model\label{app_bun}}
For unweighted undirected networks, each graph $\mathbf{G}$ in the ensemble is uniquely specified by its binary symmetric adjacency matrix $\mathbf{A}$ with entries $a_{ij}=a_{ji}=1$ if vertices $i$ and $j$ are connected, and $a_{ij}=a_{ji}=0$ otherwise. Generally, one considers loop-less graphs with $a_{ii}=0$ unless otherwise specified. This fixes the first step of our method according to the list shown above. 
Thus we can replace $\mathbf{G}\to\mathbf{A}$ and $g_{ij}\to a_{ij}$ in our general notation used so far.

Given a real binary undirected network $\mathbf{A}^*$ with entries $\{a_{ij}^*\}$ and  degree sequence $\vec{k}(\mathbf{A}^*)$, our method allows to compare the properties of $\mathbf{A}^*$ with those displayed by a randomized ensemble of binary undirected graphs having, on average, the same degree sequence as $\mathbf{A}^*$.
As we mentioned in sec. \ref{sec_limitations}, the available methods have severe limitations. 
In particular, as noted in refs.\cite{newman_origin} and \cite{mylikelihood}, the incorrectness of eq.(\ref{eq_cl})
is a consequence of the fact that it is not a proper maximum-entropy probability over the ensemble of binary graphs, i.e. it cannot be traced back to a Hamiltonian model as the ones described in \ref{app_maxent}.
By contrast, our method provides the correct solution.
The appropriate choice is to include the constraint $\vec{C}=\vec{k}$ into eq.(\ref{eq_H}) and obtain the corresponding correct probability \cite{newman_expo}.
This is precisely what the steps 2-4 of our method prescribe. For the sake of completeness, we briefly sketch the main results. If $\vec{C}(\mathbf{A})=\vec{k}(\mathbf{A})$, the Hamiltonian reads
\begin{equation}
H(\mathbf{A},\vec{\theta})=\sum_i \theta_i k_i(\mathbf{A})=\sum_{i<j}(\theta_i+\theta_j)a_{ij}
\end{equation}
The partition function can be calculated exactly \cite{newman_expo} as
\begin{equation}
Z(\vec{\theta})=\sum_\mathbf{A}e^{-H(\mathbf{A},\vec{\theta})}=\prod_{i<j}(1+e^{-\theta_i-\theta_j})
\end{equation}
Therefore the graph probability can be written in the factorized form (\ref{eq_factorize}) as follows
\begin{equation}
P(\mathbf{A}|\vec{\theta})
=\prod_{i<j}D_{ij}(a_{ij},a_{ji}|\vec{\theta})
=\prod_{i<j}P_{ij}(a_{ij}|\vec{\theta})
\end{equation}
where 
\begin{equation}
P_{ij}(a_{ij}|\vec{\theta})=p_{ij}^{a_{ij}}(1-p_{ij})^{(1-a_{ij})}
\end{equation}
is the \emph{mass probability function} of a Bernoulli-distributed binary random variable $a_{ij}$, with success probability 
\begin{equation}
p_{ij}=\frac{e^{-\theta_i-\theta_j}}{1+e^{-\theta_i-\theta_j}}
\end{equation}
representing the probability that a link between $i$ and $j$ is present.
Introducing the new variable $x_i\equiv e^{-\theta_i}$, not to be confused with the symbol $X$ used so far, and changing notation from $\vec{\theta}$ to $\vec{x}$, the expectation value of $a_{ij}$ is simply given by
\begin{equation}
\langle a_{ij}\rangle_{\vec{x}}=p_{ij}=\frac{x_ix_j}{1+x_ix_j}
\label{eq_bun_moment1}
\end{equation}
Also, since $a_{ij}^2=a_{ij}$, the second moment is
\begin{equation}
\langle a_{ij}^2\rangle_{\vec{x}}=\langle a_{ij}\rangle_{\vec{x}}
\label{eq_bun_moment2}
\end{equation}
Finally, if $(i,j)$ and $(t,s)$ are two distinct pairs of vertices, then $a_{ij}$ and $a_{ts}$ are independent random variables and 
\begin{equation}
\langle a_{ij}a_{ts}\rangle_{\vec{x}}=\langle a_{ij}\rangle_{\vec{x}}\langle a_{ts}\rangle_{\vec{x}}
\end{equation}
This completes the fourth step in our method.

The fifth step consists in finding the particular parameter values $\vec{x}^*$ that maximize eq.(\ref{eq_likelihoodlocal}), that in this case reads
\begin{equation}
\mathcal{L}(\vec{x})\equiv \ln P({\mathbf{A}^*}|\vec{x})=\sum_i k_i(\mathbf{A}^*)\ln x_i-\sum_{i<j}\ln (1+x_i x_j)
\label{eq_likelihoodbun}
\end{equation}
Equivalently \cite{mylikelihood}, the parameters $\vec{x}^*$ can be found solving the following $N$ coupled equations enforcing the desired constraints as in eq.(\ref{eq_caveragelocal}):
\begin{equation}
\sum_{j\ne i}\frac{x^*_ix^*_j}{1+x^*_ix^*_j}=k_i(\mathbf{A}^*) \qquad \forall i
\end{equation}
Importantly, since $x_i\equiv e^{-\theta_i}$ and $\theta_i$ is a real number, the solution we are looking for is the one where $x^*_i>0$ $\forall i$. This solution is unique.
Even for large networks, the above parameter estimation ranges from seconds to tens of seconds even on an ordinary laptop. 

Once the parameters $\vec{x}^*$ are found, we can proceed to the sixth step and exploit eq.(\ref{eq_expXgeneral}) to obtain the expectation values of the properties $X$ of interest:
\begin{equation}
\langle X\rangle^*=
\sum_\mathbf{A}X(\mathbf{A})P(\mathbf{A}|\vec{x}^*)=X(\langle \mathbf{A}\rangle^*)+\dots
\label{eq_X*bun}
\end{equation}
In particular, the expectation value of the ANND defined in eq.(\ref{eq_knn}) is 
\begin{equation}
\langle k^{nn}_i\rangle^*=
\frac{\sum_{j\ne i}\sum_{k\ne j} \langle a_{ij}\rangle^*\langle a_{jk}\rangle^*}
{\sum_{j\ne i} \langle a_{ij}\rangle^*}
\label{eq_knnexp}
\end{equation}
and the expectation value of the clustering coefficient defined in eq.(\ref{eq_c}) is
\begin{equation}
\langle c_i\rangle^* =
\frac{\sum_{j\ne i}\sum_{k\ne i,j}\langle a_{ij}\rangle^*\langle a_{jk}\rangle^*\langle a_{ki}\rangle^*}
{\sum_{j\ne i}\sum_{k\ne i,j}\langle a_{ij}\rangle^*\langle a_{ki}\rangle^*}
\label{eq_cexp}
\end{equation}
where $\langle a_{ij}\rangle^*=x_i^* x_j^*/(1+x_i^* x_j^*)$.
Similarly, the standard deviation $\sigma^*[X]$ can be evaluated using eq.(\ref{eq_independentpropagation}), which here reads
\begin{equation}
\sigma^*[X]=\sqrt{
\sum_{i,j}\left(\sigma^*[a_{ij}]\frac{\partial X}{\partial a_{ij}}\right)_{\mathbf{A}=\langle\mathbf{A}\rangle^*}^2+\dots}
\label{eq_bunpropagation}
\end{equation}
where $\sigma^*[a_{ij}]=\sqrt{\langle a_{ij}\rangle^*(1-\langle a_{ij}\rangle^*)}=\sqrt{x_i^* x_j^*}/(1+x_i^* x_j^*)$.
It is straightforward to obtain $\sigma^*[X]$ in terms of $\vec{x}^*$ alone, by using the derivation rule (\ref{eq_deltarule}):
\begin{equation}
\frac{\partial a_{ij}}{\partial a_{ts}}=\delta_{it}\delta_{js}
\label{eq_deltarulebun}
\end{equation}
This can also be implemented symbolically in adequate softwares. Let us calculate explicitly the standard deviations of the constraints:
\begin{equation}
\sigma^*[k_{i}]=\sqrt{\sum_{j\neq i}(\sigma^*[a_{ij}])^2}=\sqrt{\sum_{j\neq i}p_{ij}^*(1-p_{ij}^*)}
\label{eq_degreeserror}
\end{equation}
which in turn imply that
\begin{equation}
\frac{\sigma^*[k_{i}]}{k_{i}}=\sqrt{\frac{1}{k_{i}}-\frac{\sum_{j\neq i}(p_{ij}^*)^2}{k_{i}^2}}.
\label{eq_degreeserrorfrac}
\end{equation}
Given the vertex $i$, if $p_{ij}^*\ll 1,\:j=1\dots N$ and $j\neq i$, the trend decreases as $k_{i}^{-1/2}$ (which also represents an upper-bound for the ratio). The more this condition is violated (the vertex $i$ has an high degree, there are hubs in the network, etc.), the more important becomes the correction, lowering the ratio to eventually reach zero.

\subsection{Directed configuration model\label{app_bdn}}
Binary directed networks have an asymmetric adjacency matrix $\mathbf{A}$ with entries $a_{ij}=1$ if a directed link from $i$ to $j$ is there, and $a_{ij}=0$ otherwise. 
Given a real binary directed network $\mathbf{A}^*$ with out-degree sequence $\vec{k}^{out}(\mathbf{A}^*)$ and in-degree sequence $\vec{k}^{in}(\mathbf{A}^*)$, our method provides analytical  expressions for the expectation values and standard deviations of topological properties across the maximum-entropy ensemble of binary directed graphs with out-degree sequence $\vec{k}^{out}(\mathbf{A}^*)$ and in-degree sequence $\vec{k}^{in}(\mathbf{A}^*)$.
The Hamiltonian is now
\begin{eqnarray}
H(\mathbf{A},\vec{\alpha},\vec{\beta})&=&\sum_i [\alpha_i k_i^{out}(\mathbf{A})+\beta_i k_i^{in}(\mathbf{A})]\nonumber\\
&=&\sum_{i\ne j}(\alpha_i+\beta_j)a_{ij}
\end{eqnarray}
The partition function can be calculated exactly \cite{newman_expo} as
\begin{equation}
Z(\vec{\alpha},\vec{\beta})=\sum_\mathbf{A}e^{-H(\mathbf{A},\vec{\alpha},\vec{\beta})}=\prod_{i\ne j}(1+e^{-\alpha_i-\beta_j})
\end{equation}
The graph probability is now
\begin{equation}
P(\mathbf{A}|\vec{\alpha},\vec{\beta})
=\prod_{i<j}D_{ij}(a_{ij},a_{ji}|\vec{\alpha},\vec{\beta})
=\prod_{i\ne j}P_{ij}(a_{ij}|\vec{\alpha},\vec{\beta})
\end{equation}
where 
\begin{equation}
P_{ij}(a_{ij}|\vec{\alpha},\vec{\beta})=p_{ij}^{a_{ij}}(1-p_{ij})^{(1-a_{ij})}
\end{equation}
and
\begin{equation}
p_{ij}=\frac{e^{-\alpha_i-\beta_j}}{1+e^{-\alpha_i-\beta_j}}
\end{equation}
Setting $x_i\equiv e^{-\alpha_i}$ and $y_i\equiv e^{-\beta_i}$, the expectation value of $a_{ij}$ is
\begin{equation}
\langle a_{ij}\rangle_{\vec{x},\vec{y}}=p_{ij}=\frac{x_iy_j}{1+x_iy_j}
\label{eq_bdn_moment1}
\end{equation}
The second moment is
\begin{equation}
\langle a_{ij}^2\rangle_{\vec{x},\vec{y}}=\langle a_{ij}\rangle_{\vec{x},\vec{y}}
\label{eq_bdn_moment2}
\end{equation}
Finally, if $(i,j)$ and $(t,s)$ are two distinct pairs of vertices, now including the case $(t,s)=(j,i)$, then \begin{equation}
\langle a_{ij}a_{ts}\rangle_{\vec{x},\vec{y}}=\langle a_{ij}\rangle_{\vec{x},\vec{y}}\langle a_{ts}\rangle_{\vec{x},\vec{y}}
\end{equation}
The log-likelihood (\ref{eq_likelihoodlocal}) to maximize is \begin{eqnarray}
\mathcal{L}(\vec{x},\vec{y})&=&\sum_i\left[ k^{out}_i(\mathbf{A}^*)\ln x_i+
k^{in}_i(\mathbf{A}^*)\ln y_i\right]\nonumber\\
&-&\sum_{i\ne j}\ln (1+x_i y_j)
\label{eq_likelihoodbdn}
\end{eqnarray}
and the values $\vec{x}^*$, $\vec{y}^*$ that realize the maximum can alternatively be found by solving the $2N$ coupled equations
\begin{eqnarray}
\sum_{j\ne i}\frac{x^*_iy^*_j}{1+x^*_iy^*_j}&=&k^{out}_i(\mathbf{A}^*)\qquad \forall i\\
\sum_{j\ne i}\frac{x^*_jy^*_i}{1+x^*_jy^*_i}&=&k^{in}_i(\mathbf{A}^*)\qquad \forall i
\end{eqnarray}
corresponding to eq.(\ref{eq_caveragelocal}).
Again, we are looking for the solution where $x^*_i>0$ and $y^*_i>0$ $\forall i$. 
Expectation values can still be obtained using eq.(\ref{eq_X*bun}). In particular, the directed ANNDs defined in eqs.(\ref{eq_knnout}) and (\ref{eq_knnin}) have expectation values 
\begin{eqnarray}
\langle k^{nn,out}_i\rangle^*&=&
\frac{\sum_{j\ne i}\sum_{k\ne j} \langle a_{ij}\rangle^*\langle a_{jk}\rangle^*}
{\sum_{j\ne i}\langle  a_{ij}\rangle^*}\\
\langle k^{nn,in}_i\rangle^*&=&
\frac{\sum_{j\ne i}\sum_{k\ne j} \langle a_{ji}\rangle^*\langle a_{kj}\rangle^*}
{\sum_{j\ne i}\langle  a_{ji}\rangle^*}
\end{eqnarray}
where $\langle a_{ij}\rangle^*=x_i^* y_j^*/(1+x_i^* y_j^*)$.
Similarly, the standard deviation $\sigma^*[X]$ can still be evaluated through eqs.(\ref{eq_bunpropagation}) and (\ref{eq_deltarulebun}), now using $\sigma^*[a_{ij}]=\sqrt{x_i^* y_j^*}/(1+x_i^* y_j^*)$. Let us calculate explicitly the standard deviations of the constraints:
\begin{eqnarray}
\sigma^*[k_{i}^{out}]&=&\sqrt{\sum_{j\neq i}p_{ij}^*(1-p_{ij}^*)}\\
\sigma^*[k_{i}^{in}]&=&\sqrt{\sum_{j\neq i}p_{ji}^*(1-p_{ji}^*)}
\label{eq_dirdegreeserror}
\end{eqnarray}
which in turn imply that
\begin{eqnarray}
\frac{\sigma^*[k_{i}^{out}]}{k_{i}^{out}}&=&\sqrt{\frac{1}{k_{i}^{out}}-\frac{\sum_{j\neq i}(p_{ij}^*)^2}{(k_{i}^{out})^2}}\\
\frac{\sigma^*[k_{i}^{in}]}{k_{i}^{in}}&=&\sqrt{\frac{1}{k_{i}^{in}}-\frac{\sum_{j\neq i}(p_{ji}^*)^2}{(k_{i}^{in})^2}}.
\label{eq_dirdegreeserrorfrac}
\end{eqnarray}
Given the vertex $i$, if $p_{ij}^*\ll 1,\:j=1\dots N$ and $j\neq i$, the trend decreases as $(k_{i}^{out})^{-1/2}$ (which also represents an upper-bound for the ratio). The more this condition is violated (the vertex $i$ has an high out-degree, there are in-degree hubs in the network, etc.), the more important becomes the correction, lowering the ratio to eventually reach zero. Similar observations hold for the in-degrees.

\subsection{Weighted configuration model\label{app_wun}}
When weighted undirected networks are considered, each graph $\mathbf{G}$ in the ensemble is specified by its non-negative symmetric matrix $\mathbf{W}$ whose  integer entry $w_{ij}$ represents the weight of the link between vertices $i$ and $j$ (including  $w_{ij}=0$ if no link is there). 
Thus we can replace $\mathbf{G}\to\mathbf{W}$ and $g_{ij}\to w_{ij}$ in the general notation. 
As we mentioned in the main text, in the weighted configuration model a real weighted undirected network $\mathbf{W}^*$ with entries $\{w_{ij}^*\}$ is compared with a maximum-entropy ensemble of weighted  undirected graphs having the same strength sequence $\vec{s}(\mathbf{W}^*)$.
In our method, by setting $\vec{C}=\vec{s}$ into eq.(\ref{eq_H}) we obtain the Hamiltonian 
\begin{equation}
H(\mathbf{W},\vec{\theta})=\sum_i \theta_i s_i(\mathbf{W})=\sum_{i<j}(\theta_i+\theta_j)w_{ij}
\end{equation}
The partition function is \cite{mybosefermi}
\begin{equation}
Z(\vec{\theta})=\sum_\mathbf{W}e^{-H(\mathbf{W},\vec{\theta})}=\prod_{i<j}\frac{1}{1-e^{-\theta_i-\theta_j}}
\end{equation}
and is only defined if $\theta_i>0$ $\forall i$.
The graph probability is \cite{mybosefermi}
\begin{equation}
P(\mathbf{W}|\vec{\theta})
=\prod_{i<j}D_{ij}(w_{ij},w_{ji}|\vec{\theta})
=\prod_{i<j}P_{ij}(w_{ij}|\vec{\theta})
\end{equation}
where 
\begin{equation}
P_{ij}(w_{ij}|\vec{\theta})=p_{ij}^{w_{ij}}(1-p_{ij})
\end{equation}
is the mass probability function of a geometrically-distributed \cite{myWRG} integer random variable $w_{ij}$, with success probability 
\begin{equation}
p_{ij}=e^{-\theta_i-\theta_j}
\end{equation}
representing the probability that a link between $i$ and $j$ is present.
Introducing $x_i\equiv e^{-\theta_i}\in[0,1)$, the expectation value of $w_{ij}$ is 
\begin{equation}
\langle w_{ij}\rangle_{\vec{x}}=\frac{p_{ij}}{1-p_{ij}}=\frac{x_ix_j}{1-x_ix_j}
\label{eq_wun_moment1}
\end{equation}
Now in general $w_{ij}^2\ne w_{ij}$, and the second moment is
\begin{equation}
\langle w_{ij}^2\rangle_{\vec{x}}=
\frac{p_{ij}(1+p_{ij})}{(1-p_{ij})^2}=
\frac{(x_ix_j)(1+x_ix_j)}{(1-x_ix_j)^2}
\label{eq_wun_moment2}
\end{equation}
Finally, if $(i,j)$ and $(t,s)$ are two distinct pairs of vertices, then  
\begin{equation}
\langle w_{ij}w_{ts}\rangle_{\vec{x}}=\langle w_{ij}\rangle_{\vec{x}}\langle w_{ts}\rangle_{\vec{x}}
\end{equation}
The log-likelihood (\ref{eq_likelihoodlocal}) reads
\begin{equation}
\mathcal{L}(\vec{x})\equiv \ln P({\mathbf{W}^*}|\vec{x})=\sum_i s_i(\mathbf{A}^*)\ln x_i+\sum_{i<j}\ln (1-x_i x_j)
\label{eq_likelihoodwun}
\end{equation}
and the parameters $\vec{x}^*$ maximizing it solve the following $N$ coupled equations 
\begin{equation}
\sum_{j\ne i}\frac{x^*_ix^*_j}{1-x^*_ix^*_j}=s_i(\mathbf{A}^*) \qquad \forall i
\end{equation}
enforcing the desired constraints as in eq.(\ref{eq_caveragelocal}).
Now the solution must be looked for in the region $0\le x_i<1$ $\forall i$.

Through the parameters $\vec{x}^*$ we obtain the expectation values of the properties $X$ of interest:
\begin{equation}
\langle X\rangle^*=
\sum_\mathbf{W}X(\mathbf{W})P(\mathbf{W}|\vec{x}^*)=X(\langle \mathbf{W}\rangle^*)+\dots
\label{eq_X*wun}
\end{equation}
For instance, the expectation value of the weighted ANND defined in eq.(\ref{eq_knnw}) is 
\begin{equation}
\langle\tilde{k}^{nn}_i\rangle^*=
\frac{\sum_{j\ne i}\sum_{k\ne j} \langle w_{ij}\rangle^*\langle w_{jk}\rangle^*}
{W^*\sum_{j\ne i} \langle w_{ij}\rangle^*}
\end{equation}
where we have used $\langle W\rangle^*=W*$ (see main text).
Similarly, the weighted clustering coefficient defined in eq.(\ref{eq_cw}) has expectation value
\begin{equation}
\langle \tilde{c}_i\rangle^*=
\frac{\sum_{j\ne i}\sum_{k\ne i,j}\langle w_{ij}\rangle^*\langle w_{jk}\rangle^*\langle w_{ki}\rangle^*}
{W^*\sum_{j\ne i}\sum_{k\ne i,j}\langle w_{ij}\rangle^*\langle w_{ki}\rangle^*}
\end{equation}
where $\langle w_{ij}\rangle^*=x_i^* x_j^*/(1-x_i^* x_j^*)$.
Similarly, according to eq.(\ref{eq_independentpropagation}) the standard deviation $\sigma^*[X]$ is
\begin{equation}
\sigma^*[X]=\sqrt{
\sum_{i,j}\left(\sigma^*[w_{ij}]\frac{\partial X}{\partial w_{ij}}\right)_{\mathbf{W}=\langle\mathbf{W}\rangle^*}^2+\dots}
\label{eq_wunpropagation}
\end{equation}
where $\sigma^*[w_{ij}]=\sqrt{x_i^* x_j^*}/(1-x_i^* x_j^*)$.
The rule (\ref{eq_deltarule}) here reads
\begin{equation}
\frac{\partial w_{ij}}{\partial w_{ts}}=\delta_{it}\delta_{js}
\end{equation}
and allows to obtain $\sigma^*[X]$ in terms of $\vec{x}^*$ alone. Let us calculate explicitly the standard deviations of the constraints:
\begin{equation}
\sigma^*[s_{i}]=\sqrt{\sum_{j\neq i}(\sigma^*[w_{ij}])^2}=\sqrt{\sum_{j\neq i}\langle w_{ij}\rangle^*(1+\langle w_{ij}\rangle^*)}
\label{eq_wdegreeserror}
\end{equation}
which in turn imply that
\begin{equation}
\frac{\sigma^*[s_{i}]}{s_{i}}=\sqrt{\frac{1}{s_{i}}+\frac{\sum_{j\neq i}(\langle w_{ij}\rangle^*)^2}{s_{i}^2}}.
\label{eq_wdegreeserrorfrac}
\end{equation}
Given the vertex $i$, if $\langle w_{ij}\rangle^*\ll 1,\:j=1\dots N$ and $j\neq i$, the trend decreases as $s_{i}^{-1/2}$. The more this condition is violated (the vertex $i$ has an high strength, there are `strength-hubs' in the network, etc.), the more important becomes the correction. Note that for weighted networks the second term has a positive sign. This means that the correction `increases' the $s_{i}^{-1/2}$ trend which now represents a lower-bound for the coefficient of variation.

\section{\uppercase{Nonlocal constraints}}
Our model can also be applied to more complicated cases where the constraints are no longer local. 
However, a necessary condition for our method to work with nonlocal constraints is that eq.(\ref{eq_X2}) can still be expressed exactly in a form which does not require the enumeration of all possible graphs (in other words, the partition function can be calculated analytically). In such a case, eq.(\ref{eq_caverage}) can still be used to calculate the parameters $\vec{\theta}^*$ exactly as in the local case, and at the same time those parameters can be used to obtain the analytical expressions for the expected value and standard deviation of the topological properties of interest.
Therefore, only a limited number of nonlocal constraints lend themselves to an analytical treatment. However, since the philosophy of randomization algorithms is always to enforce the simplest constraints in order to detect higher-order patterns, it turns out that the mathematically tractable constraints are also the ones of major interest. 
We now provide an explicit example of a choice of nonlocal constraints that is often used in empirical studies, and at the same time preserves the analytical character of our method and yields exact results.

\subsection{Reciprocal configuration model\label{app_rdn}}
As discussed in the main text, a more constrained null model for a binary directed network $\mathbf{A}^*$ is one where the three reciprocal degree sequences $\vec{k}^\rightarrow(\mathbf{A}^*)$, $\vec{k}^\leftarrow(\mathbf{A}^*)$ and $\vec{k}^\leftrightarrow(\mathbf{A}^*)$ are specified,
where 
\begin{eqnarray}
k_i^\rightarrow(\mathbf{A}^*)&\equiv&\sum_{j\ne i}a^*_{ij}(1-a^*_{ji})\\
k_i^\leftarrow(\mathbf{A}^*)&\equiv&\sum_{j\ne i}a^*_{ji}(1-a^*_{ij})\\
k_i^\leftrightarrow(\mathbf{A}^*)&\equiv&\sum_{j\ne i}a^*_{ij}a^*_{ji}
\end{eqnarray}
The Hamiltonian for this model is
\begin{equation}
H(\mathbf{A},\vec{\alpha},\vec{\beta},\vec{\gamma})=\sum_i [\alpha_i k_i^\rightarrow(\mathbf{A})+\beta_i k_i^\leftarrow(\mathbf{A})+\gamma_i k_i^\leftrightarrow(\mathbf{A})]
\nonumber
\end{equation}
The nonlocality is manifest in the fact that, unlike the previous examples, now the (second-order) constraints involve products of two adjacency matrix entries.
Despite this complication, the partition function can still be calculated exactly \cite{mygrandcanonical} as
\begin{equation}
Z(\vec{\alpha},\vec{\beta},\vec{\gamma})=\prod_{i< j}(1+e^{-\alpha_i-\beta_j}+e^{-\alpha_j-\beta_i}+e^{-\gamma_i-\gamma_j})
\end{equation}
The graph probability can still be expressed in the form (\ref{eq_factorize}), i.e. 
\begin{equation}
P(\mathbf{A}|\vec{\alpha},\vec{\beta},\vec{\gamma})
=\prod_{i<j}D_{ij}(a_{ij},a_{ji}|\vec{\alpha},\vec{\beta},\vec{\gamma})
\end{equation}
In the above expression, 
\begin{equation}
D_{ij}(a_{ij},a_{ji}|\vec{\alpha},\vec{\beta},\vec{\gamma})=
(p^\rightarrow_{ij})^{a^\rightarrow_{ij}}(p^\leftarrow_{ij})^{a^\leftarrow_{ij}}
(p^\leftrightarrow_{ij})^{a^\leftrightarrow_{ij}}
(p^\nleftrightarrow_{ij})^{a^\nleftrightarrow_{ij}}
\nonumber
\end{equation}
is the dyadic probability defined in terms of
\begin{eqnarray}
a_{ij}^\rightarrow&\equiv& a_{ij}(1-a_{ji})\\
a_{ij}^\leftarrow&\equiv& a_{ji}(1-a_{ij})\\
a_{ij}^\leftrightarrow&\equiv& a_{ij}a_{ji}\\ a_{ij}^\nleftrightarrow&\equiv& (1-a_{ij})(1-a_{ji})
\end{eqnarray}
and 
\begin{eqnarray}
p_{ij}^\rightarrow&\equiv&\langle a_{ij}^\rightarrow\rangle_{\vec{x},\vec{y},\vec{z}}=\frac{x_i y_j}{1+x_i y_j+x_j y_i +z_i z_j}
\label{eq_exprecipriniz}\\
p_{ij}^\leftarrow&\equiv& \langle a_{ij}^\leftarrow\rangle_{\vec{x},\vec{y},\vec{z}}=\frac{x_j y_i}{1+x_i y_j+x_j y_i +z_i z_j}\\
p_{ij}^\leftrightarrow&\equiv&\langle a_{ij}^\leftrightarrow\rangle_{\vec{x},\vec{y},\vec{z}}=\frac{z_i z_j}{1+x_i y_j+x_j y_i +z_i z_j}\\
p_{ij}^\nleftrightarrow&\equiv&\langle a_{ij}^\nleftrightarrow\rangle_{\vec{x},\vec{y},\vec{z}}=\frac{1}{1+x_i y_j+x_j y_i +z_i z_j}
\label{eq_expreciprfine}
\end{eqnarray} 
where we have set $x_i\equiv e^{-\alpha_i}$, $y_i\equiv e^{-\beta_i}$ and $z_i\equiv e^{-\gamma_i}$ \cite{mybosefermi}.
The above expressions represent the dyadic expectation values. 

A little algebra leads to the log-likelihood 
\begin{eqnarray}
&\mathcal{L}(\vec{x},\vec{y},\vec{z})=
-\sum_{i< j}\ln (1+x_i y_j+x_j y_i+z_i z_j)+&\nonumber\\
&\sum_i\left[ k^\rightarrow_i(\mathbf{A}^*)\ln x_i+
k^\leftarrow_i(\mathbf{A}^*)\ln y_i+
k^\leftrightarrow_i(\mathbf{A}^*)\ln z_i\right]&\nonumber\\
\nonumber
\end{eqnarray}
and the values $\vec{x}^*$, $\vec{y}^*$, $\vec{z}^*$ that realize the maximum can alternatively \cite{mylikelihood} found by solving the $3N$ coupled equations
\begin{eqnarray}
\sum_{j\ne i}\frac{x^*_i y^*_j}{1+x^*_i y^*_j+x^*_j y^*_i+ z_i^* z_j^*}
&=&k^\rightarrow_i(\mathbf{A}^*)\qquad \forall i\nonumber\\
\sum_{j\ne i}\frac{x^*_j y^*_i}{1+x^*_i y^*_j+x^*_j y^*_i+ z_i^* z_j^*}
&=&k^\leftarrow_i(\mathbf{A}^*)\qquad \forall i\nonumber\\
\sum_{j\ne i}\frac{z^*_i z^*_j}{1+x^*_i y^*_j+x^*_j y^*_i+ z_i^* z_j^*}
&=&k^\leftrightarrow_i(\mathbf{A}^*)\qquad \forall i\nonumber\\
\nonumber
\end{eqnarray}
corresponding to an example when eq.(\ref{eq_caverage}) can be written explicitly even if the constraints are nonlocal.
We are looking for the solution where $x^*_i>0$, $y^*_i>0$ and $z^*_i>0$ $\forall i$. 

The expectation values of topological properties involving products of dyadic terms can be obtained exactly without resorting to the linear approximation in eq.(\ref{eq_expXgeneral}).
For instance, the number of occurrences of a particular motif $m$, where $m$ labels one of the possible 13 non-isomorphic connected motifs with three vertices, is 
\begin{equation}
N_m\equiv\sum_{i\ne j\ne k}a^{m,1}_{ij}a^{m,2}_{jk}a^{m,3}_{ki}
\end{equation}
where $a^{m,l}_{ij}$ is one of the four possible dyadic relations $a^\rightarrow_{ij}$, $a^\leftarrow_{ij}$, $a^\leftrightarrow_{ij}$, $a^\nleftrightarrow_{ij}$, and 
$\{a^{m,1}_{ij},a^{m,2}_{jk},a^{m,3}_{ki}\}$ indicates the specific triplet of dyadic relations defining motif $m$.
The exact expectation value of $N_m$ is
\begin{equation}
\langle N_m\rangle^*\equiv\sum_{i\ne j\ne k}\langle a^{m,1}_{ij}\rangle^* \langle a^{m,2}_{jk}\rangle^* \langle a^{m,3}_{ki}\rangle^* 
\end{equation}
where $\langle a^{m,1}_{ij}\rangle^*$ is given by evaluating eqs.(\ref{eq_exprecipriniz})-(\ref{eq_expreciprfine}) at the particular values $\vec{x}^*$, $\vec{y}^*$, $\vec{z}^*$.
The standard deviation of $N_m$, and in general of a topological property $X$, can still be obtained using eq.(\ref{eq_independentpropagation}), i.e.
\begin{eqnarray}
(\sigma^*[X])^2&=&
\sum_{i,j}\left[\left(\sigma^*[a_{ij}]\frac{\partial X}{\partial a_{ij}}\right)_{\mathbf{A}=\langle\mathbf{A}\rangle^*}^2\right.\\
&+&\left.\sigma^*[a_{ij},a_{ji}]\left(\frac{\partial X}{\partial a_{ij}}\frac{\partial X}{\partial a_{ji}}\right)_{\mathbf{A}=\langle\mathbf{A}\rangle^*}\right]
+\dots\nonumber
\end{eqnarray}
where now
\begin{eqnarray}
(\sigma^*[a_{ij}])^2&=&\langle a_{ij}\rangle^*
(1-\langle a_{ij}\rangle^*)\nonumber\\
&=&\langle a^\leftrightarrow_{ij}+a^\rightarrow_{ij}\rangle^*
(1-\langle a^\leftrightarrow_{ij}+a^\rightarrow_{ij}\rangle^*)\nonumber
\end{eqnarray}
and
\begin{eqnarray}
\sigma^*[a_{ij},a_{ji}]&=&\langle a_{ij}a_{ji}\rangle^*-\langle a_{ij}\rangle^*\langle a_{ji}\rangle^*\nonumber\\
&=&\langle a_{ij}^\leftrightarrow\rangle^*-\langle a^\leftrightarrow_{ij}+a^\rightarrow_{ij}\rangle^*
\langle a^\leftrightarrow_{ji}+a^\rightarrow_{ji}\rangle^*
\nonumber
\end{eqnarray}
which are both known exactly in terms of eqs.(\ref{eq_exprecipriniz})-(\ref{eq_expreciprfine}). The calculations for the standard deviations of the constraints are similar to the directed configuration model case:
\begin{equation}
\sigma^*[k_{i}^{a}]=\sqrt{\sum_{j\neq i}(p_{ij}^{a})^*(1-(p_{ij}^{a})^*)}
\label{eq_recdegreeserror}
\end{equation}
which in turn imply that
\begin{equation}
\frac{\sigma^*[k_{i}^{a}]}{k_{i}^{a}}=\sqrt{\frac{1}{k_{i}^{a}}-\frac{\sum_{j\neq i}((p_{ij}^a)^*)^2}{(k_{i}^{a})^2}}
\label{eq_recdegreeserrorfrac}
\end{equation}
(where $a=\rightarrow,\:\leftarrow,\:\leftrightarrow$) and similar observations hold.

\section{\uppercase{Comparison with computational microcanonical algorithms}\label{app_comparison}}
\begin{figure}[t!]
\begin{center}
\includegraphics[width=.48\textwidth]{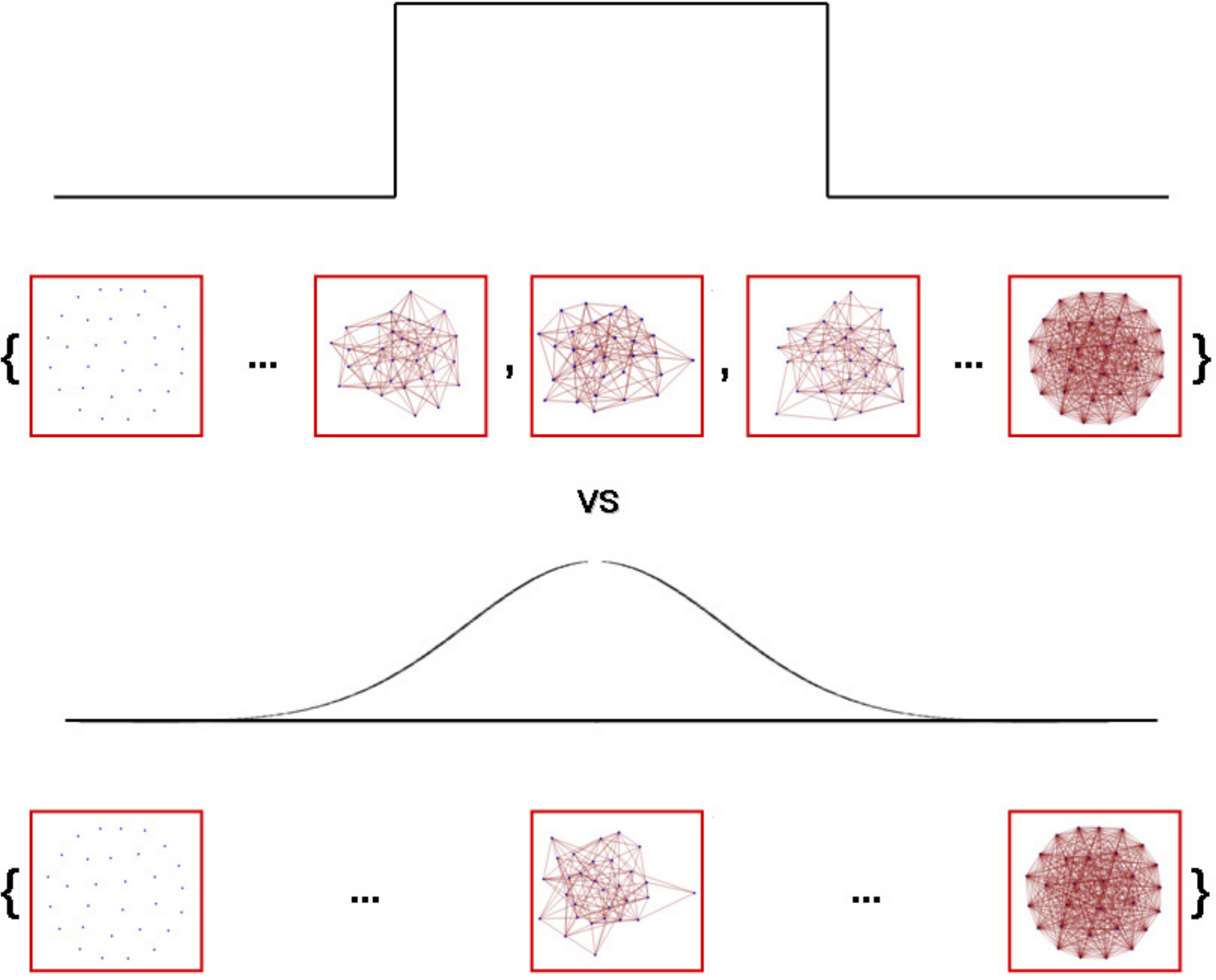}
\end{center}
\caption{Difference between the LRA-based microcanonical approach and our likelihood-based grandcanonical approach. Top: the microcanonical approach assigns non-zero probability only to the subset $\mathcal{D}(\vec{C})$ of graphs that realize the enforced constraints $\vec{C}$ (in the example shown, a given value of the number of links $L$) exactly.
Bottom: by contrast, our grandcanonical approach assigns non-zero probability to all graphs, but this probability reaches its maximum value for the graphs belonging to $\mathcal{D}(\vec{C})$. In so doing, it is more robust to potential errors in the original network data (such as missing links).
\label{fig_distrib}}
\end{figure}

The LRA-based microcanonical approach  \cite{MS,MSZ} and our likelihood-based grandcanonical approach are in general not equivalent for finite networks. Let $\mathcal{D}(\vec{C})$ be the set of all graphs $\mathbf{G}$ that realize the enforced constraints $\vec{C}=\{C_\alpha\}$ exactly. Both approaches assign equal probabilities to all graphs that realize the constraints, i.e. $P(\mathbf{G}_1)=P(\mathbf{G}_2)$ if $\mathbf{G}_1\in \mathcal{D}(\vec{C})$ and $\mathbf{G}_2\in \mathcal{D}(\vec{C})$.
Also, in both approaches these graphs are the most likely to occur, i.e. $P(\mathbf{G}_1)>P(\mathbf{G}_2)$ for any $\mathbf{G}_1\in \mathcal{D}(\vec{C})$ and $\mathbf{G}_2\notin \mathcal{D}(\vec{C})$. However the two approaches are different, the microcanonical one being very severe in assigning zero probability to any graph where the degrees are not matched exactly, i.e. $P(\mathbf{G})=0$ if $\mathbf{G}\notin \mathcal{D}(\vec{C})$. By contrast, in the grandcanonical approach all possible graphs can occur, even if with very different probabilities, in such a way that the ensemble average of the desired constraints over all graphs coincides with the observed values (see fig.\ref{fig_distrib} for an illustration of this difference). 

The above key and elegant property places grandcanonical ensembles at the basis of information theory. Notably, they are more robust to errors in the original data such as missing or overrepresented links. In presence of even a small percentage of such errors, the `true' graph (the unobserved one affected by errors) will never appear in the microcanonical ensemble, while it will appear with nonzero probability in the grandcanonical ensemble. As desirable, for small deviations from the observed graph the true graph will have a slightly decreased probability with respect to the one assigned by our method to the observed graph, while for larger errors the probabilities will differ by a larger amount.

Therefore, while for infinite systems the microcanonical and grandcanonical ensembles become equivalent since fluctuations about the average values become negligible, in finite systems the use of grandcanonical ensembles is preferable. What is of interest for us here is the impact of the two methods on the topological properties induced on the randomized networks. To this end, we now show explicitly the relation between the two approaches when applied to particular networks. We shall only consider unweighted networks for simplicity.

In the unweighted (either directed or undirected) case, our method directly provides `from the beginning' the explicit values of the probabilities $p^G_{ij}$ that a link from $i$ to $j$ is there. The superscript $G$ stands for `grandcanonical', and the probability is evaluated at the parameter values that maximize the likelihood, as described above. By contrast, the microcanonical approach samples the configuration space iteratively, and the microcanonical probability $p^M_{ij}$ that a link from $i$ to $j$ is there can only be evaluated as the frequency of occurrence of the link over many randomizations. As the number of randomized networks increases, this frequency will converge to $p^M_{ij}$. However this asymptotic value will also depend on the number $R$ of elementary rewiring steps used to obtain a single randomized network. 
\begin{figure}[b!]
\begin{center}
\includegraphics[width=.48\textwidth]{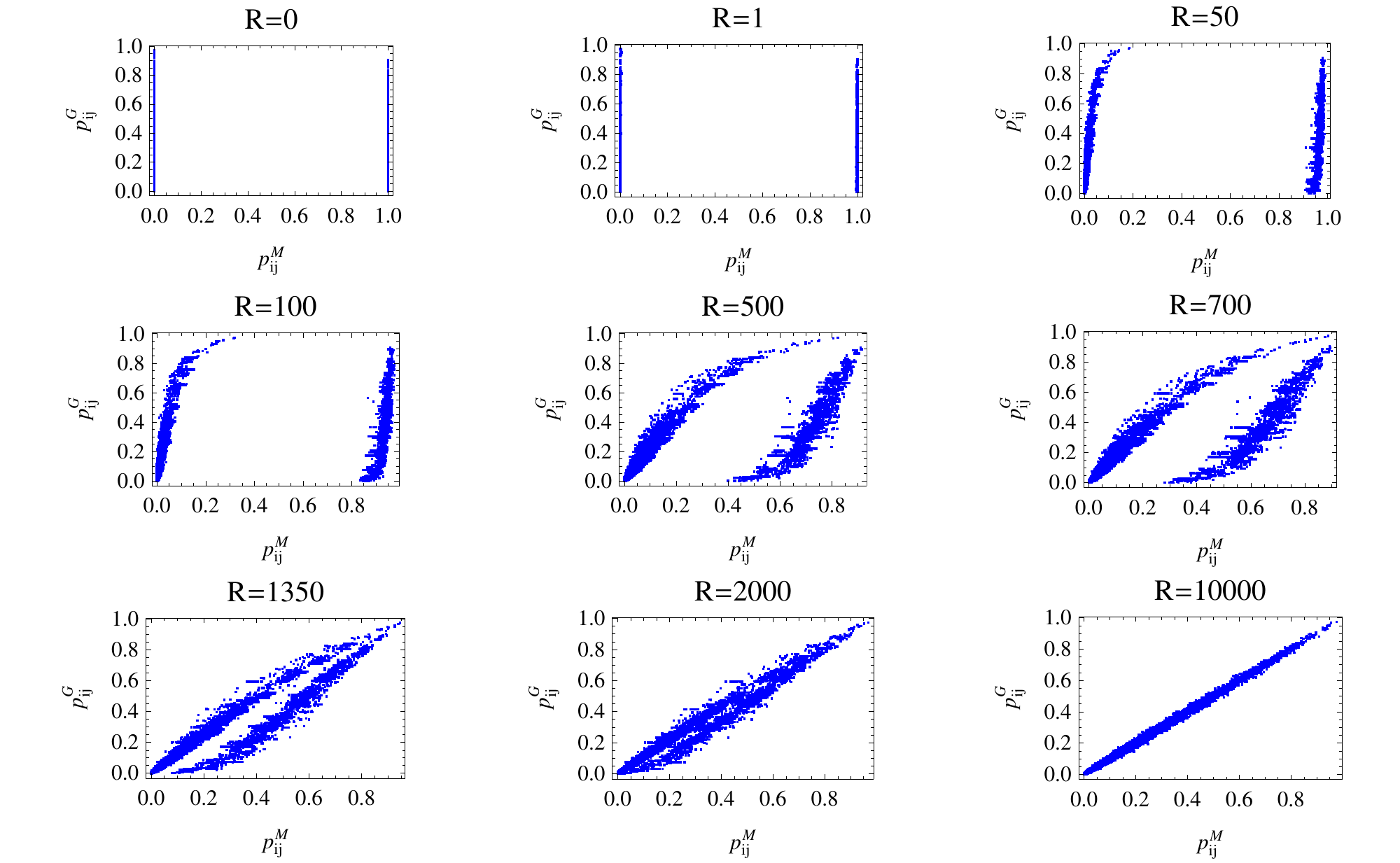}
\end{center}
\caption{Convergence of the microcanonical connection probability $p^M_{ij}$ (measured using the local rewiring algorithm) to the grandcanonical probability $p^G_{ij}$ (obtained using our maximum-entropy method) as the number $R$ of local rewiring moves per network increases.\label{fig_comparison}}
\end{figure}
To see this, consider the trivial case $R=0$. As no rewiring takes place, all the `randomized' networks will in fact coincide with the original network. If the adjacency matrix of the latter has elements $\{a_{ij}\}$, this means that $p^M_{ij}=a_{ij}$. If $R$ is nonzero but still very small, $p^M_{ij}$ will not change substantially. Only if $R$ is large enough then $p^M_{ij}$ will approach $p^G_{ij}$. This is shown explicitly in fig.\ref{fig_comparison}, where we plot $p^G_{ij}$ as a function of $p^M_{ij}$ for all directed pairs of vertices $(i,j)$ by taking the \emph{Little Rock Lake} food web as the starting network. As $R$ increases from $R=0$ to $R=10000$, the double-peaked shape (corresponding to $p^M_{ij}=a_{ij}$ independently of $p^G_{ij}$) evolves towards the identity $p^M_{ij}=p^G_{ij}$. Similar evolution patterns are observed for all the networks we analyzed. This clearly shows that in our method we obtain `from the beginning' the values $p^G_{ij}$ to which the microcanonical $p^M_{ij}$ will converge only after several iterations. Notably, the number $R$ of rewiring steps required for $p^M_{ij}$ to converge to $p^G_{ij}$ acceptably is not known \emph{a priori} and without the knowledge of $p^G_{ij}$ itself. This problematic aspect of the microcanonical approach highlights another advantage of the grandcanonical one.

The two approaches are in general not equivalent for finite networks. We can now state this more rigorously, and indicate at least two ways in which they may differ. 

First of all, $p^M_{ij}$ represent marginal probabilities, where the information about the correlations between the presence of a link between different pairs of vertices has been lost. While in the grandcanonical approach these correlations are absent, and different pairs of vertices are always statistically independent, in the microcanonical approach some weak correlations will be preserved even after many rewiring steps. These correlations arise from the microcanonical constraint of matching the degree sequence (or other contraints) exactly. Thus, while our grandcanonical method enables to compute the expected topological properties exactly, in the microcanonical approach this is not possible.

Secondly, the final `convergence' of $p^M_{ij}$ to $p^G_{ij}$ for $R\to\infty$ will in general not hold exactly. This means that the asymptotic plot of $p^G_{ij}$ versus $p^M_{ij}$ will not be a strict identity, but a narrow scatter of points close to the identity. In other words, increasing $R$ beyond a certain value will not make the quantities converge further. For some networks (such as the \emph{Little Rock Lake} food web shown above) one may attain a better convergence than for others.

It is interesting to understand whether the degree of convergence between the two approaches depends on some property of the network.
To this end, we first define two measures of discrepancy between $\{p^G_{ij}\}$ and $\{p^M_{ij}\}$, and then study how they behave on well-controlled, artificially generated networks.
As measures of discrepancy, we consider the $l^2$ distance
\begin{equation}
\Delta_{l^2}\equiv\sqrt{
\frac{\sum_{i\ne j}|p^G_{ij}-p^M_{ij}|^2}
{N(N-1)}
}
\label{eq_l2}
\end{equation}
and the Kullback-Leibler information distance
\begin{eqnarray}
&\Delta_{KL}\equiv\displaystyle{\frac{\sum_{i\ne j}p^M_{ij}(\log_2 p^M_{ij}-\log_2 p^G_{ij})}{N(N-1)}}+&\\
&\displaystyle{\frac{\sum_{i\ne j}(1-p^M_{ij})[\log_2 (1-p^M_{ij})-\log_2 (1-p^G_{ij})]}{N(N-1)}}&\nonumber
\label{eq_KL}
\end{eqnarray}

\noindent (note that we have normalized the above distances in such a way that both lie in the range $[0,1]$). It is instructive to use these distances to compare the two methods on a family of artificially generated networks. We considered $N=100$ vertices, assigned each vertex a real value $x_i$ drawn randomly in the interval $[0,1]$, and established an edge between each pair of vertices $i$ and $j$ with probability $p_{ij}=z x_i x_j/(1+z x_i x_j)$. 

This choice generates maximally random networks with degree distribution controlled by $\{x_i\}$ as in eq.(\ref{eq_bun_moment1}), but has an additional parameter $z$ that tunes the overall link density $d\equiv 2L/N(N-1)$, representing the fraction of realized links. With $\{x_i\}$ kept constant, we considered various choices of $z$ and, for each of them, adopted both the microcanonical randomization and our grandcanonical method. 

\begin{figure}[h!]
\begin{center}
\includegraphics[width=.48\textwidth]{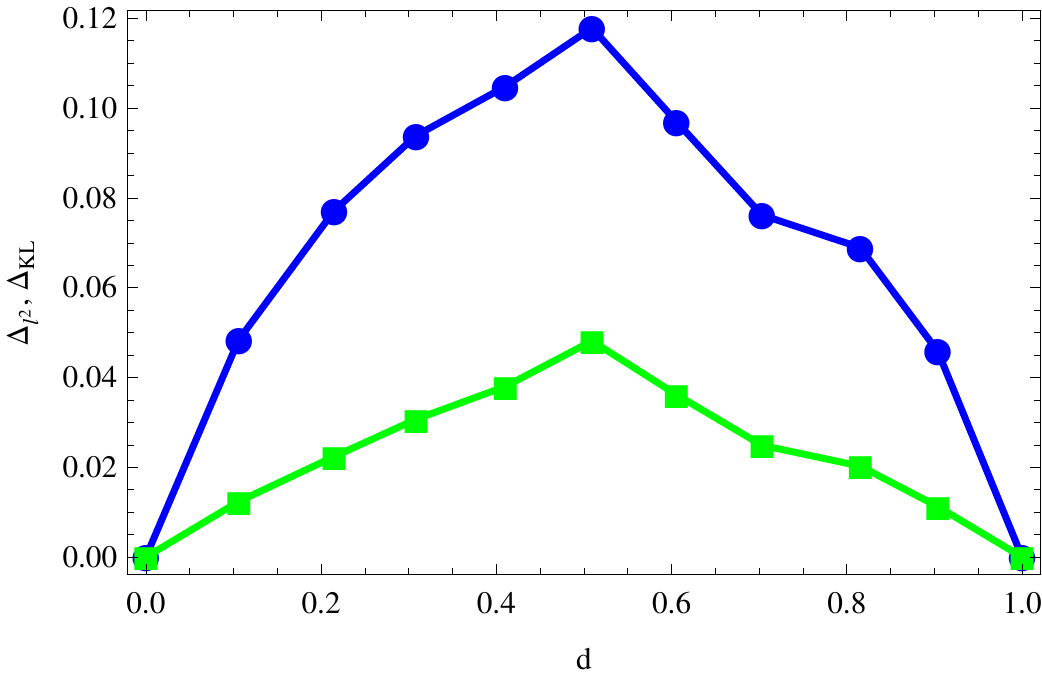}
\end{center}
\caption{Kullback-Leibler ($\Delta_{KL}$, green squares) and $l^2$ ($\Delta_{l^2}$, blue circles) distance between microcanonical ($p_{ij}^M$) and grandcanonical ($p_{ij}^G$) marginal connection probabilities, plotted versus link density $d$.\label{fig_distances}}
\end{figure}

In fig.\ref{fig_distances} we show the resulting difference between the marginal probabilities $\{p^G_{ij}\}$ and $\{p^M_{ij}\}$, as a function of link density. The two methods yield very similar results for both small and large link density, whereas for intermediate density values they display a greater difference. Even in this case, however, the distances between them are $\Delta_{l^2}\approx 0.05$ and $\Delta_{KL}\approx 0.12$, both small considering their possible range of variation.
\begin{acknowledgments}
D.G. acknowledges financial support from the European Commission 6th FP (Contract CIT3-CT-2005-513396), Project: DIME - Dynamics of Institutions and Markets in Europe.
\end{acknowledgments}

\end{document}